\documentclass[%
 aip,
 amsmath,amssymb,
 reprint,%
]{revtex4-1}

\usepackage{xcolor}
\usepackage{graphicx}% Include figure files
\usepackage{dcolumn}% Align table columns on decimal point
\usepackage{bm}% bold math
\usepackage[utf8]{inputenc}
\usepackage{mathptmx}
\usepackage{etoolbox}
\usepackage{subfigure}
\makeatletter
\def\@email#1#2{%
 \endgroup
 \patchcmd{\titleblock@produce}
  {\frontmatter@RRAPformat}
  {\frontmatter@RRAPformat{\produce@RRAP{*#1\href{mailto:#2}{#2}}}\frontmatter@RRAPformat}
  {}{}
}%
\makeatother
\begin{document}
\preprint{AIP/123-QED}

\title[Preprint]{Phase-Field Simulations for Dripping-to-Jetting Transitions:
Effects of Low Interfacial Tension and Bulk Diffusion }

% Force line breaks with \\
\author{Fukeng Huang}
%\email{hfkeng@nus.edu.sg}
 %\altaffiliation%{Department of Mathematics, National University of Singapore, Singapore, 119076}%Lines break automatically or can be forced with \\

\author{Weizhu Bao }%
 %\email{matbaowz@nus.edu.sg}
\affiliation{Department of Mathematics, National University of Singapore, Singapore, 119076}%

\author{Tiezheng Qian }
\homepage{Corresponding author: maqian@ust.hk}
%\email{Corresponding author: maqian@ust.hk}
\affiliation{ Department of Mathematics, The Hong Kong University of Science and Technology\\
 Clear Water Bay, Kowloon, Hong Kong, P. R. China
}%

%\date{\today}% It is always \today, today,
             %  but any date may be explicitly specified

\begin{abstract}
The dripping-to-jetting transitions in coaxial flows have been experimentally well studied
for systems of high interfacial tension,
where the capillary number of the outer fluid and the Weber number of the inner fluid are in control.
Recent experiments have shown that in systems of low interfacial tension, the transitions driven by the inner flow
are no longer dominated by the inertial force alone, and the viscous drag force due to the inner flow is also
quantitatively important.
In the present work, we carry out numerical simulations based on the Cahn-Hilliard-Navier-Stokes model,
aiming for a more complete and quantitative study to understand
the effects of interfacial tension when it becomes sufficiently low.
The Cahn-Hilliard-Navier-Stokes model is solved by using an accurate and efficient spectral method
in a cylindrical domain with axisymmetry.
%%%%%%%%%
Plenty of numerical examples are systematically presented to show the dripping-to-jetting transitions
driven by the outer flow and inner flow respectively.
In particular, for transitions dominated by inner flow, detailed results reveal
how the magnitude of interfacial tension quantitatively determines the relative importance of
the inertial and viscous forces due to the inner flow at the transition point.
Our numerical results are found to be consistent with the experimental observation.
Finally, the degree of bulk diffusion is varied to investigate its quantitative effect on
the condition for the occurrence of transition. Such effect is expected for systems
of ultralow interfacial tension where interfacial motion is more likely to be driven by bulk diffusion. 
\end{abstract}

\maketitle

\section{\label{sec:intro} Introduction}

Dripping and jetting in coaxial flows of two immiscible fluids refer to the phenomena in which two fluids are forced to flow through
a cylindrical conduit, with one fluid, namely the inner fluid, flowing at the center and the other, namely the outer fluid,
flowing around it in a coaxial manner.
When the flow rates of both fluids are low, dripping occurs due to the capillary instability, with the inner fluid forming
discrete drops close to the orifice. On the other hand, when the flow rates are high enough,
jetting occurs, with the inner fluid forming a continuous jet
that extends out of the orifice and breaks into drops further downstream.
The dripping and jetting of coaxial flows of two immiscible fluids have many applications
in ink jet printing, biomedical engineering, and materials engineering \cite{whitesides2006origins,stone2004engineering},
and the transition from dripping to jetting is of fundamental importance in these applications involving drop formation \cite{marre2009dripping,kaufman2012structured}.
Extensive research efforts have been made to study the dripping-to-jetting transition in coaxial geometry
\cite{utada2007dripping,guillot2007stability,castro2009scaling}.
Among these works, Utada {\it et al.} \cite{utada2007dripping} demonstrated that the transitions in coflowing streams
can be characterized by the capillary number of the outer fluid and the Weber number of the inner fluid.
The dripping-to-jetting transitions have also been investigated in other geometries such as flow-focusing
\cite{ganan1998generation,cubaud2008capillary} and T-junction \cite{thorsen2001dynamic,abate2009impact}.
For a review summarizing the main observations and understandings for common device geometries, we refer to \cite{nunes2013dripping} and the references therein.
%%%%%%%%%

% previous works on high interfacial tension and our aims on low interfacial tension
To understand the hydrodynamics of the dripping-to-jetting transitions, most of the previous studies
have focused on systems, e.g., oil-water ones, of high interfacial tension. For these systems,
as demonstrated in \cite{utada2007dripping}, the dripping-to-jetting transitions can be described by a state diagram
that is controlled by the capillary number of the outer fluid and the Weber number of the inner fluid.
This means that for transitions driven by the inner fluid, the effect of the viscous force due to the inner flow is negligible.
However, it has been shown experimentally in \cite{mak2017dripping} that when the interfacial tension is sufficiently low,
the dripping-to-jetting transitions driven by the inner fluid are no longer dominated by the inertial force alone,
and the viscous force due to the inner flow also plays a quantitatively important role.
Therefore, for a comprehensive understanding of the dripping-to-jetting transitions,
a more complete and quantitative study is needed to investigate the effect of interfacial tension when it becomes sufficiently low.
This will help clarify the relative importance of the inertial and viscous forces due to the inner flow
in inducing the transitions.
In addition, recent observations in aqueous two-phase systems with ultra-low surface tension have revealed novel and interesting
pinch-off dynamics dominated by bulk diffusion \cite{Xu2019}.
To the best of our knowledge, the effect of bulk diffusion on the dripping-to-jetting transitions in systems of ultra-low
interfacial tension has never been investigated before.
%%%%%%%%%

The main purpose of the present work is to investigate the dripping-to-jetting transitions in coaxial flows
over a wide range of interfacial tension and with variable bulk diffusion.
Firstly, we aim to numerically observe and examine the dripping-to-jetting transitions driven by the outer and inner fluids respectively.
This is to provide a crucial indicator to distinguish jetting from dripping and hence locate the point of transition
and the critical flow rate.
Secondly, regarding the contributions of inertial and viscous forces due to the inner flow,
we aim to establish a quantitative relationship between them at the point of transition over a wide range of interfacial tension.
Last but not least, we will examine the quantitative effect of bulk diffusion on the critical flow rate at the point of transition.
This will also show if bulk diffusion can change the relative importance of the inertial and viscous forces due to the inner flow
at the transition point.
%%%%%%%%%

% previous work on numerical simulations and our method

While numerous experimental studies having been carried out on the dripping-to-jetting transitions in coaxial flows,
there have been very few works focusing on numerical simulations.
Guillaument {\it et al.} \cite{guillaument2013numerical} utilized the one-fluid model and the volume of fluid method
to simulate segmented micro coflows of CO$_2$ and water in two dimensions.
Lei {\it et al.} \cite{lei2011dripping} employed the phase-field model to investigate two types of transitions
driven by the outer flow and the inner flow in two dimensions.
Shahin {\it et al.} \cite{shahin2017three} simulated dripping and jetting in a coflowing system using
a one-fluid model in three dimensions and developed a novel algorithm to handle the topological change of the interface mesh.

To investigate the dripping-to-jetting transitions in immiscible two-phase flows,
we will employ the Cahn-Hilliard-Navier-Stokes (CHNS) model and carry out numerical computation
in a cylindrical domain with axisymmetry.
The phase-field methods have been widely used in the simulations of interfacial motion in multiphase flows
\cite{anderson1998diffuse,boettinger2002phase,yue2004diffuse,Qian2003,Qian2004,yang2006JCP}
as they avoid the need of interface tracking and can easily and efficiently accommodate topological changes such as pinch-off,
a key feature of the dripping and jetting phenomena.

To the best of our knowledge, there has been no prior work that investigates the dripping-to-jetting transitions
in three dimensions using the phase-field method.
For the CHNS model employed here, a characteristic length scale
has been introduced in \cite{huang2022diffuse} to measure the competition between diffusion and viscous flow in interfacial motion.
Parameters involved in defining this length scale can be adjusted to tune the effect of bulk diffusion in the simulated system.
Numerically, we solve the CHNS model by using the spectral method \cite{Shen97} for the spatial discretization
and the pressure-correction method \cite{shen1992error,lopez1998efficient} for the temporal discretization.
These methods have been demonstrated to be accurate and efficient in treating the phase-field models in cylindrical domains \cite{yang2006JCP,lopez1998efficient,lopez2002efficient}.

% organization
This paper is organized as follows.
In Sec. \ref{sec:CHNS}, the CHNS model is derived by applying Onsager's variational principle \cite{Qian2006,Onsager1931a,onsager1931b}.
In Sec. \ref{sec:dimensionless}, the dimensionless equation system is presented with important dimensionless parameters
associated with the dripping-to-jetting transitions, and the simulated systems are described in a cylindrical domain
with necessary boundary conditions for the inner and outer flows with adjustable flow rates.
In Sec. \ref{sec:results_discussion}, numerical results are presented to show the distinct between dripping and jetting
in the regime dominated by the outer flow and that by the inner flow, respectively.
Furthermore, in the regime dominated by the inner flow, the relative importance of the inertial and viscous forces
at the transition point is investigated over a wide range of interfacial tension,
with numerical results showing agreement with recent experiments.
Finally, the quantitative effect of bulk diffusion on the critical flow rates at the transition point is also measured.
In Sec. \ref{sec:conclusion}, the paper is concluded with a few remarks.
%%%%%%%%%

\section{\label{sec:model} Modeling and simulation for immiscible two-phase flows}

\subsection{\label{sec:CHNS} The Cahn-Hilliard-Navier-Stokes model}

Consider a multi-component fluid with two co-existing immiscible phases. A diffuse-interface model uses
a Ginzberg-Landau-type free energy functional to describe the thermodynamic properties of the fluid.
Here we use the Cahn-Hilliard (CH) free energy functional \cite{CH1958}
\begin{equation}\label{eq:CH-free-energy}
F_{\rm CH}[\phi]=\int \left[\displaystyle\frac{K}{2}\left(\nabla\phi \right)^2 +f(\phi) \right] d\mathbf{r},
\end{equation}
in which $\phi:=\phi(\mathbf{r})$ is the phase-field variable to measure the local relative concentration,
$f(\phi)$ is the Helmholtz free energy density for a homogeneous phase, and $K$ is a positive material parameter.
The free energy density $f$ is given by $f(\phi)=-\frac{\alpha}{2}\phi^2 + \frac{\beta}{4}\phi^4$,
which has a double-well structure to stabilize the fluid-fluid interface between the two co-existing phases around
$\phi_{\pm}=\pm \phi_0=\pm \sqrt{ \frac{\alpha}{\beta} }$,
where $\alpha$ and $\beta$ are two positive parameters.
Subject to appropriate boundary conditions, $F_{\rm CH}[\phi]$ can be minimized
to stabilize a flat interface between the two equilibrium phases of $\phi=\pm \phi_0$.
The interfacial structure gives the interfacial tension $\gamma=\frac{2\sqrt{2}\alpha^2\xi}{3\beta}$
and the characteristic length scale $\xi=\sqrt{\frac{K}{\alpha}}$ for the interfacial thickness \cite{Qian2006}.
Note that in many literatures, $\phi_0$ is made to equal $1$ through a rescaling. Here
$\phi_0=\sqrt{ \frac{\alpha}{\beta} }$ is purposely retained to measure the distance away from
the critical point where $\phi_0$ vanishes.
%%%%%%%%%

For an incompressible fluid, the velocity field $\mathbf{v}$ is subject to the incompressibility condition
$\nabla\cdot \mathbf{v}=0$, and the phase field $\phi$ satisfies the continuity equation
\begin{equation}\label{eq:continuity-phi}
\displaystyle\frac{\partial \phi}{\partial t}=-\nabla\cdot\mathbf{J}
=-\nabla\cdot\left( \phi\mathbf{v} + \mathbf{j} \right),
\end{equation}
where $\mathbf{J}=\mathbf{\phi\mathbf{v} + \mathbf{j}}$ is the total current density, in which
$\phi\mathbf{v}$ is contributed by the flow and $\mathbf{j}$ is the diffusive current density
contributed by the bulk diffusion.

Hydrodynamic equations for immiscible two-phase flows can be derived by
applying Onsager's variational principle (cf. appendix A in \cite{huang2022diffuse}) as follows.
The Rayleighian ${\cal {R}}$ is given by ${\cal R}=\dot{F}_{\rm CH}+\Phi$ in the bulk region.
Here $\dot{F}_{\rm CH}$ is the rate of change of $F_{\rm CH}[\phi]$, given by
\begin{equation}\label{eq:CH-free-energy-rate}
\dot{F}_{\rm CH}[\phi]=\int \mu\displaystyle\frac{\partial \phi}{\partial t} d\mathbf{r}
=\int  \nabla\mu\cdot\left( \mathbf{\phi\mathbf{v} + \mathbf{j}} \right) d\mathbf{r},
\end{equation}
in which $\mu = \frac{\delta F_{\rm CH} }{\delta\phi}$ is the chemical potential, given by
$\mu=-K\nabla^2\phi + f'(\phi)$, and the continuity equation (\ref{eq:continuity-phi}) has been used
with the impermeability conditions for $\mathbf{v}$ and $\mathbf{j}$ at the solid boundary.
The other part in ${\cal R}$ is the dissipation functional $\Phi$, which is half the rate of free energy dissipation
and given by
\begin{equation}\label{eq:dissipation-functional-B}
\Phi=\int  \displaystyle\frac{\eta}{4}
\left[ \nabla\mathbf{v} + \left( \nabla\mathbf{v} \right)^T \right]^2 d\mathbf{r}
+ \int  \displaystyle\frac{\mathbf{j}^2}{2M} d\mathbf{r},
\end{equation}
which is contributed by the viscous dissipation, with $\eta$ being the shear viscosity,
and the diffusive dissipation, with $M$ being the mobility coefficient.

Subject to the incompressibility condition, the Rayleighian can be minimized with respect to the rates
$\mathbf{v}$ and $\mathbf{j}$. This gives the force balance equation
\begin{equation}\label{eq:force-balance-equation}
-\nabla p + \nabla\cdot \bm{\sigma}_{\rm visc} - \phi\nabla \mu=0
\end{equation}
for $\mathbf{v}$,
and the constitutive equation
\begin{equation}\label{eq:diffusive-current-density}
\mathbf{j}=-M\nabla\mu
\end{equation}
for $\mathbf{j}$.
Here $p$ is the pressure, which is the Lagrange multiplier to locally impose $\nabla\cdot \mathbf{v}=0$,
$\bm{\sigma}_{\rm visc}$ is the Newtonian stress tensor given by
$\bm{\sigma}_{\rm visc}=\eta \left[ \nabla\mathbf{v} + \left( \nabla\mathbf{v} \right)^T \right]$.
Equation (\ref{eq:force-balance-equation}) is the Stokes equation with the capillary force density, and
it can be readily generalized to the Navier-Stokes equation
\begin{equation}\label{eq:NS-equation}
\rho\left[ \displaystyle\frac{\partial \mathbf{v}}{\partial t} +
\left(\mathbf{v}\cdot\nabla\right) \mathbf{v}\right]=
-\nabla p + \nabla\cdot \bm{\sigma}_{\rm visc} - \phi\nabla \mu.
\end{equation}
Combining equations (\ref{eq:continuity-phi}) and (\ref{eq:diffusive-current-density}) gives
the advection-diffusion equation for the phase field $\phi$:
\begin{equation}\label{eq:CH-equation}
\displaystyle\frac{\partial \phi}{\partial t} + \mathbf{v}\cdot \nabla \phi
=-\nabla\cdot \mathbf{j} = M \nabla^2 \mu,
\end{equation}
which is the CH equation for a constant mobility $M$.
Equations (\ref{eq:NS-equation}) and (\ref{eq:CH-equation}) govern the hydrodynamics of immiscible two-phase flows.
In the present work, the simplest situation is treated with the two fluids having 
equal density, equal viscosity and equal mobility. 

\subsection{\label{sec:dimensionless} Dimensionless equations and simulated systems}

Numerical simulations are carried out by solving the CHNS system:
\begin{subequations}
\begin{align}
& \frac{\partial \phi}{\partial t}+\mathbf{v} \cdot \nabla \phi=M \nabla^2 \mu,\\
& \mu=-K\nabla^2 \phi-\alpha \phi+\beta \phi^3,\\
& \rho \big(\frac{\partial \mathbf{v} }{\partial t}+\mathbf{v} \cdot \nabla \mathbf{v}  \big)=-\nabla p+\eta \nabla^2 \mathbf{v} +\mu \nabla \phi, \label{eq:CHNS3}\\
& \nabla \cdot \mathbf{v} =0,
\end{align}
\end{subequations}
in a cylindrical domain $\Omega=\{\mathbf{r}=(x,y,z): x^2+y^2 <L^2, z \in (0,H) \}$.
Here $M$, $K$, $\alpha$, $\beta$, $\rho$, and $\eta$ are material parameters introduced in Sec. \ref{sec:CHNS}.
Note that the pressure $p$ in equation \eqref{eq:CHNS3} is different from that in equation \eqref{eq:NS-equation},
with $-\phi \nabla \mu$ there being replaced by $\mu \nabla \phi$ here.
The boundary conditions on $x^2+y^2=L^2$ are
\begin{equation}
%\begin{align}
\frac{\partial \phi}{\partial \mathbf{n}}=0, \quad
\frac{\partial \mu}{\partial \mathbf{n}}=0, \quad
\mathbf{v} =0.
%\end{align}
\end{equation}
In our simulations, two immiscible phases flow into the cylinder on the boundary $z=0$ and
out of the cylinder on the boundary $z=H$. The boundary conditions there for $\phi$ and $\mu$ are given by
\begin{equation}
\phi=\tanh \Big(\frac{r-R}{\sqrt{2}\xi} \Big), \quad
\mu=0,
\end{equation}
on $z=0$, with $r=\sqrt{x^2+y^2}$, $R$ being the radius of the inner tube, and
\begin{equation}
\frac{\partial \phi}{\partial \mathbf{n}}=0, \quad
\frac{\partial \mu}{\partial \mathbf{n}}=0,
\end{equation}
on $z=H$.
%%%
Finally, the boundary conditions for $\mathbf{v}:=(v_x,v_y,v_z)$ on $z=0$ and $z=H$ are given by
\begin{equation}
v_x=v_y=0,\quad v_z=\left\{
\begin{array}{lr}
a(R^2-r^2),\quad 0<r<R,\\
-b(r^2-R^2)+\frac{b(L^2-R^2)}{\ln\frac{L}{R}}\ln\frac{r}{R},\quad  R\le r< L,
\end{array}
\right. %\quad \text{on}\quad z=0,%
\end{equation}
on $z=0$, where $a$ and $b$ are the parameters determining
the mean velocities (i.e., flow rates) of the inner and outer phases, respectively, and
\begin{equation}
v_x=v_y=0,\quad v_z=c(L^2-r^2), %\quad \text{on}\quad z=H,%
\end{equation}
on $z=H$, where $c$ is the parameter determining the mean velocity of the flow out of the cylinder.
Note that these flow profiles are based on the Poiseuille profile, and the parameters $a$, $b$ and $c$ satisfy
\begin{equation}
aR^4 + b(L^4-R^4) - b\frac{(L^2-R^2)^2}{\ln\frac{L}{R}}=cL^4,
\end{equation}
for the volume conservation.

To nondimensionalize the above system, we use the radius $L$ of the computational domain $\Omega$ as
the length unit, $u=\frac{\gamma}{\eta}$ as the velocity unit, $\tau=\frac{L}{u}$ as the time unit,
and $p_0=\frac{\eta}{\tau}$ as the pressure unit.
We also define the following quantities:
\begin{itemize}
\item  $\bar{H}=\frac{H}{L}$ as the dimensionless length of the computational domain,
\item  $\bar{R}=\frac{R}{L}$ as the dimensionless radius of the inner tube,
\item $\phi_0=\sqrt{\frac{\alpha}{\beta}}$, with the two equilibrium phases separated by a flat interface
being of $\phi=\pm \phi_0$,
\item  $\varepsilon=\frac{\xi}{L} =\frac{1}{L}\sqrt{\frac{K}{\alpha}}$ as the dimensionless interfacial thickness
of the diffuse interface,
\item $D=2 M \alpha$ as the diffusion coefficient for $\phi$ close to $\pm\phi_0$ far away from the interface,
\item $l_c=\frac{\sqrt{M\eta}}{\phi_0}$ as the characteristic length scale,
determined from the competition between diffusion and viscous flow \cite{huang2022diffuse},
\item $\gamma=\frac{2\sqrt{2}}{3}\alpha \phi_0^2 {\xi}$ as the interfacial tension,
\item $Re_{\gamma}=\frac{\rho uL}{\eta}$ as the Reynolds number defined from
the velocity unit $u=\frac{\gamma}{\eta}$ and the length unit $L$,
\item $B=\frac{\eta D}{\alpha \phi_0^2 L^2}=\frac{2 l_c^2}{L^2}$ as the dimensionless parameter measuring
the characteristic length scale $l_c$ with respect to $L$.
\end{itemize}
 Dimensionless variables, denoted by using overbar, are defined as follows:
\begin{itemize}
\item
$\bar{\phi}=\frac{\phi}{\phi_0}$, \quad
$\bar{\mathbf{v}} =\frac{\mathbf{v}}{u}$,\quad
$\bar{\mu}=\frac{\mu}{\alpha \phi_0 \varepsilon}$, \quad
$\bar{p}=\frac{p}{p_0}$,
\end{itemize}
and the dimensionless operators:
\begin{itemize}
\item
 $\frac{\partial}{\partial \bar t}=\tau \frac{\partial}{\partial {t}}$, \quad $\bar \nabla =L {\nabla}$.
\end{itemize}
%%%%%%%%%
Using the above definitions, we obtain the dimensionless CHNS system in the cylindrical domain
$\bar{\Omega}=\{(\bar{x},\bar{y},\bar{z}): \bar{x}^2+\bar{y}^2 <1, \bar{z} \in (0,\bar{H}) \}$ as
\begin{small}
\begin{subequations}\label{CHNS}
\begin{align}
& \frac{\partial \bar \phi}{\partial \bar t}+\bar {\mathbf{v} } \cdot \bar{\nabla} \bar \phi=\frac{3}{4\sqrt{2}} B\bar{\nabla}^2 \bar \mu,\label{CHNS1}\\
& \bar \mu=- \varepsilon \bar \nabla^2 \bar \phi+\frac{1}{ \varepsilon }(-\bar \phi+\bar \phi^3),\label{CHNS2}\\
& Re_{\gamma}\left(
\frac{\partial \bar {\mathbf{v} }}{\partial \bar t}+\bar {\mathbf{v} } \cdot \bar \nabla \bar {\mathbf{v} } \right)
=-\bar \nabla \bar p+ \bar \nabla^2 \bar {\mathbf{v} }+\frac{3}{2\sqrt{2}}\bar \mu \bar \nabla \bar \phi,\label{CHNS3} \\
& \bar \nabla \cdot \bar {\mathbf{v} }=0.
\end{align}
\end{subequations}
\end{small}
The boundary conditions are
\begin{equation}
\frac{\partial \bar \phi}{\partial \mathbf{n}}=0, \quad
\frac{\partial \bar \mu}{\partial \mathbf{n}}=0, \quad
\mathbf{\bar v} =0,
\end{equation}
on $\bar{x}^2+\bar{y}^2=1$,
\begin{equation}
\begin{array}{l}
\bar{\mu}=0, \quad \bar{\phi}=\tanh(\frac{\bar{r}-\bar{R}}{\sqrt{2} \varepsilon}), \\
\bar{v}_x=\bar{v}_y=0, \quad
\bar{v}_z=\left\{
\begin{array}{lr}
\bar{a}(\bar{R}^2-\bar{r}^2), \quad 0<\bar{r}<\bar{R}, \\
-\bar{b}(\bar{r}^2-\bar{R}^2)+\frac{\bar{b}(1-\bar{R}^2)}{\ln\frac{1}{\bar{R}}}\ln\frac{\bar r}{\bar R}, \quad \bar{R}\le \bar{r}<1,
\end{array}
\right.
\end{array}
\end{equation}
on $\bar{z}=0$ with $\bar{r}=\sqrt{\bar{x}^2+\bar{y}^2}$, and
\begin{equation}
\begin{array}{l}
\frac{\partial \bar \phi}{\partial \mathbf{n}}=0, \quad \frac{\partial \bar \mu}{\partial \mathbf{n}}=0, \\
\bar{v}_x=\bar{v}_y=0, \quad \bar{v}_z=\bar{c}(1-\bar{r}^2),
\end{array}
\end{equation}
on $\bar{z}=\bar{H}$,
with the dimensionless parameters $\bar{a}$, $\bar{b}$ and $\bar{c}$ satisfying
\begin{equation}
\bar{a} \bar R^4+\bar{b}(1-\bar R^4)-\bar{b}\frac{(1-\bar R^2)^2}{\ln\frac{1}{\bar R}}=\bar{c}.
\end{equation}
% parameter values
{Here the dimensionless $\bar{a}$, $\bar{b}$ and $\bar{c}$ are obtained by multiplying the dimensional ones by $\frac{L^2}{u}$.}
The above dimensionless CHNS system involves the dimensionless parameters
$ \varepsilon$, $Re_{\gamma}$, $B$, $\bar{R}$, $\bar{H}$, $\bar{a}$, and $\bar{b}$.
Here $\varepsilon$ is the dimensionless interfacial thickness, which is the smallest length to be resolved,
$Re_{\gamma}$ is the Reynolds number defined from the velocity unit $u=\frac{\gamma}{\eta}$,
$B$ controls the competition between bulk diffusion and viscous flow,
$\bar R$ measures the size of the orifice (i.e., the radius of the inner tube), $\bar{H}$ measures the length of the computational domain,
and $\bar{a}$ and $\bar{b}$ control the flow rates of the inner and outer fluids.

From our three-dimensional (3D) simulations, it is verified that given an axisymmetric initial condition
in the cylindrical domain, the axisymmetry can be accurately preserved during the whole dynamic process.
Therefore, in the absence of any evidence for non-axisymmetric modes, we treat the axisymmetric 3D problem
as a reduced two-dimensional (2D) problem by making use of the cylindrical coordinates
to improve the computational efficiency \cite{huang2022diffuse}.
Technically, we first transform the 3D problem into a 2D problem 
using the cylindrical coordinates \cite{Shen97}. 
We then adopt the usual semi-implicit scheme to solve the phase-field variable 
and the spectral-projection method to solve the velocity and pressure fields 
for the Navier-Stokes equation in cylindrical geometry \cite{lopez1998efficient}. 
At each time step, we can efficiently solve a series of Poisson-type equations 
with constant coefficients.

\section{\label{sec:results_discussion} Results and discussion}
With the dimensionless parameters introduced in the previous section, the average velocity of the inner flow $\bar{v}_{\rm in}$
and that of the outer flow $\bar{v}_{\rm out}$ are given by
 \begin{equation}\label{v_in_out}
 \bar{v}_{\rm in}=\frac{1}{2}\bar{a} \bar{R}^2, \quad  \bar{v}_{\rm out}=\frac{\bar{b}}{2}(1+\bar{R}^2)+\frac{\bar{b}}{2}\frac{1-\bar{R}^2}{\ln \bar{R}}.
 \end{equation}
Using $\bar{v}_{\rm in}$ and $\bar{v}_{\rm out}$, the capillary number of the outer flow $\mathcal{C}_{\rm out}$
and the Weber number of the inner flow $\mathcal{W}_{\rm in}$ can be expressed as
\begin{equation}
\mathcal{C}_{\rm out}=\bar{v}_{\rm out}, \quad \mathcal{W}_{\rm in}=\bar{v}_{\rm in}^2 Re_{\gamma}\bar{R}.
\end{equation}
Here $\mathcal{C}_{\rm out}$ is defined by $\mathcal{C}_{\rm out} = \frac{\eta (\bar{v}_{\rm out} u)}{\gamma}$, and
$\mathcal{W}_{\rm in}$ is defined by $\mathcal{W}_{\rm in} = \frac{\rho (\bar{v}_{\rm in} u)^2 R }{\gamma} $, where
$\bar{v}_{\rm in} u$ and $\bar{v}_{\rm out} u$ are the dimensional average velocities with $u$ being the velocity unit.
%%%%%%%%%
Physically, the capillary number measures the viscous drag force, and
the Weber number measures the inertial force relative to the interfacial tension force.

It has been well established that there are two classes of dripping-to-jetting transitions in coflowing streams \cite{utada2007dripping}.
The first one is driven by strong outer flows and will be numerically investigated in Section \ref{sec:b_regime}
by fixing a small $\bar{v}_{\rm in}$ and varying $\bar{v}_{\rm out}$.
The second one is driven by strong inner flows, and will be numerically investigated in Section \ref{sec:a_dominant}
by fixing a small $\bar{v}_{\rm out}$ and varying $\bar{v}_{\rm in}$.
In this regime, our numerical results show that in addition to the inertial force measured by $\mathcal{W}_{\rm in}$,
the viscous force due to the inner flow, measured by the capillary number $\mathcal{C}_{\rm in}=\bar{v}_{\rm in}$,
also contributes to the occurrence of dripping-to-jetting transition
when the interfacial tension is sufficiently low.
This numerical observation is in agreement with recent experiments \cite{mak2017dripping}.
Finally, Section \ref{sec:B_effect} demonstrates the quantitative effect of bulk diffusion
on the critical flow rates at the transition point.
Such effect is expected for systems of ultralow interfacial tension
where interfacial motion is more likely to be driven by bulk diffusion \cite{Xu2019}.
%%%%%%%%%

\subsection{\label{sec:b_regime} Transitions dominated by outer flows}
In this subsection, we investigate the first class of dripping-to-jetting transitions driven by strong outer flows.
For this purpose, the value of $\bar{v}_{\rm in}$ is fixed to be small, and the value of $\bar{v}_{\rm out}$ is increased to
induce the transition.

\begin{figure}[!htbp]
 \centering
  \subfigure{ \includegraphics[scale=.68]{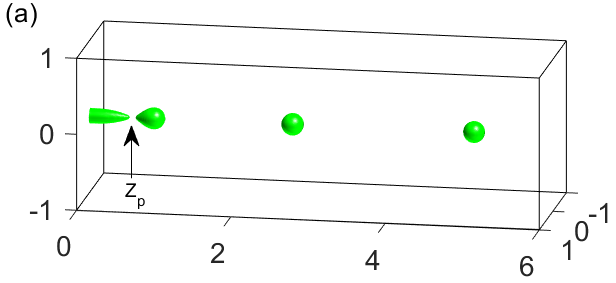}}
  \subfigure{ \includegraphics[scale=.68]{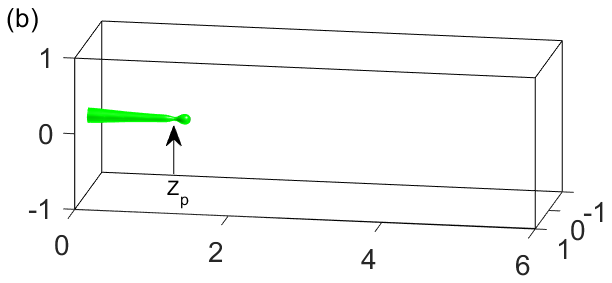}}\\
 \caption{ Two different pinch-off positions for two different outer flow rates.
 (a) A dripping state for $\bar{b}=0.575$. (b) A jetting state for $\bar{b}=0.6$.
 Other parameter values used in simulations are $\varepsilon=0.01$, $Re_{\gamma}=500$, $B=0.0002$,
 $\bar{R}=0.1$, $\bar{H}=6$, and $\bar{a}=15$.}
 \label{fig:Pb}
 \end{figure}

\begin{figure}[!htbp]
 \centering
{ \includegraphics[scale=.35]{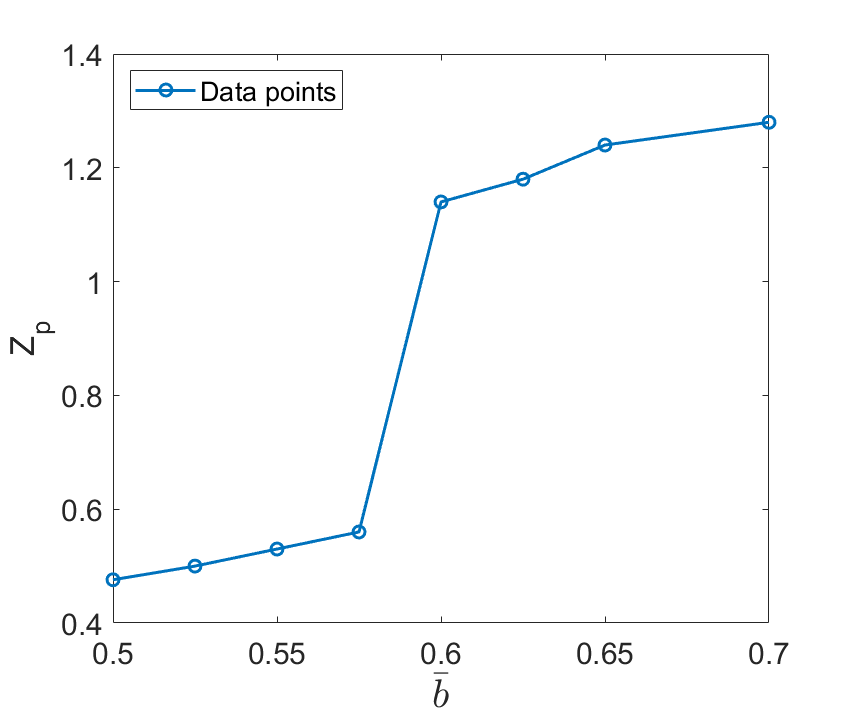}}
 \caption{ Variation of the pinch-off position $Z_p$ with the parameter $\bar{b}$ which controls the outer flow rate.
 A transition is noted to occur between $\bar{b}=0.575$ (dripping in figure \ref{fig:Pb}(a))
 and $\bar{b}=0.6$ (jetting in figure \ref{fig:Pb}(b)).
 Other parameter values used in simulations are $\varepsilon=0.01$, $Re_{\gamma}=500$, $B=0.0002$,
 $\bar{R}=0.1$, $\bar{H}=6$, and $\bar{a}=15$.}
 \label{fig:Jumpb}
 \end{figure}

We start by demonstrating the dripping-to-jetting transitions in the regime dominated by strong outer flows.
Let $Z_p$ denote the distance between the pinch-off position and the boundary of $\bar{z}=0$ (the orifice).
For a slow inner flow with $\bar{a}=15$ being fixed, $Z_p$ is expected to increase with the increasing outer flow rate, i.e.,
the increasing $\bar{b}$.
Figure \ref{fig:Pb} shows two different pinch-off positions for two different outer flow rates.
It is clearly observed from figure \ref{fig:Jumpb} that $Z_p$ exhibits a sharp increase from
$\bar b=0.575$ (figure \ref{fig:Pb}(a)) to $\bar{b}=0.6$ (figure \ref{fig:Pb}(b)),
indicating a transition from a dripping state to a jetting state
as $\mathcal{C}_{\rm out}$ is increased from $0.1450$ to $0.2030$. This critical magnitude of $\mathcal{C}_{\rm out}$
is in agreement with the experimental results in \cite{utada2007dripping}.
According to the state diagram reported in \cite{utada2007dripping}, for the dripping-to-jetting transitions dominated by outer flows, the critical values of $\mathcal{C}_{\rm out}$ are typically distributed between $0.2$ and $0.4$. 

%From Figure \ref{fig:P} and Figure \ref{fig:Pb}, we notice that either increasing only the inner flow or only the outer flow can lead the transition from dripping to jetting while the jetting phenomena are different:  The jetting driven by the strong inner flow will generate a wide jet and large drops while the jetting driven by the strong outer flow will lead a long and narrow jet and small drops.

\begin{figure}[!htbp]
 \centering
  \subfigure{ \includegraphics[scale=.5]{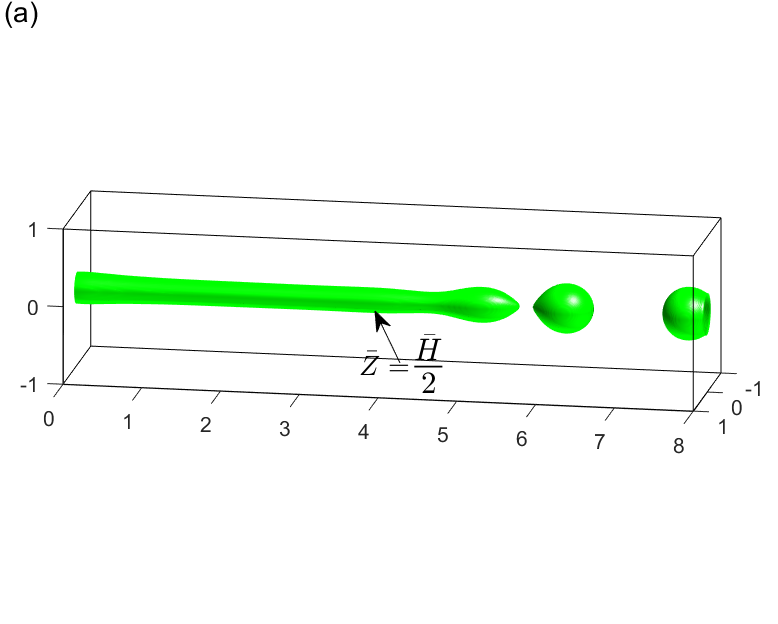}}
  \subfigure{ \includegraphics[scale=.55]{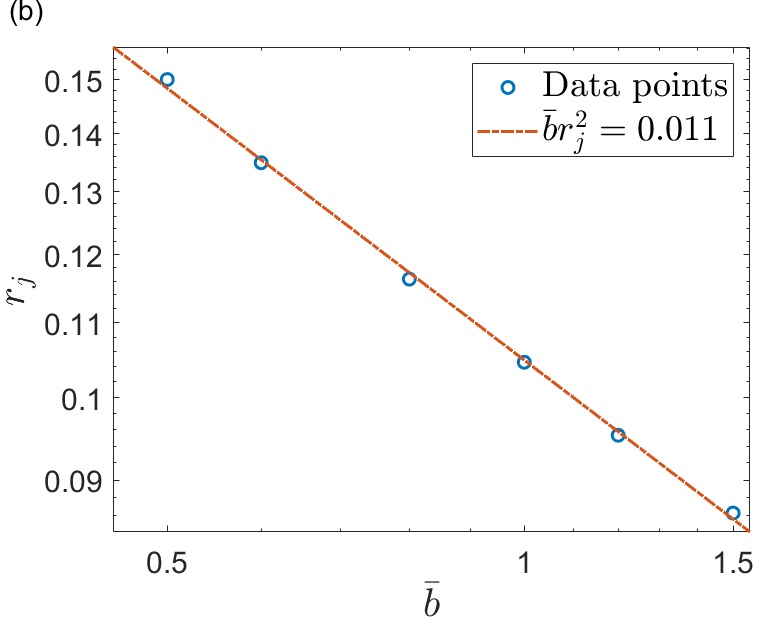}}\\
 \caption{ (a) A jetting state obtained from our simulations.
 Note that the radius of the jet $r_j$ is measured at the plane of $\bar{z}=\frac{\bar{H}}{2}$.
 (b) Log-log plot of the jet radius $r_j$ versus $\bar{b}$ which controls the outer flow rate.
 Here the mass conservation of the inner fluid is ensured by $\bar{b} r_j^2 \approx 0.011$.
 The jetting state in (a) is obtained for $\bar{b}=0.5$.
 Other parameter values used in simulations are $\varepsilon=0.01$, $Re_{\gamma}=10$, $B=0.0002$,
 $\bar{R}=0.2$, $\bar{H}=8$, and $\bar{a}=6$.}
 \label{fig:jet}
 \end{figure}

A jetting state maintained by a strong outer flow is characterized by a long, narrow jet
and small drops \cite{utada2007dripping}. In fact, a stronger outer flow results in a narrower jet and smaller drops.
Here we present some quantitative results on the relationship between the outer flow rate and the size of the corresponding jet,
i.e., the radius of the jet.
To obtain reliable data, we have ensured that the jets are long and wide enough by using values of $\bar{b}$ and $\bar{R}$
that are sufficiently large.
Figure \ref{fig:jet}(a) presents a jetting state obtained from our simulations,
and figure \ref{fig:jet}(b) shows the dependence of the jet radius $r_j$ on the outer flow rate
($\bar{v}_{\rm out}\propto \bar{b}$), with the jet radius $r_j$ being measured at the plane of $\bar{z}=\frac{\bar{H}}{2}$.
When $\bar a$ and $\bar{R}$ are both fixed, the total flux of inner fluid is given,
and a faster outer flow (with a larger $\bar{b}$) leads to a thinner jet
in which the inner fluid flows with a larger average velocity ($\propto \bar{b}$).
According to mass conservation, $\bar{b} r_j^2$ must be a constant in order to
maintain the total flux of inner fluid, as shown in figure \ref{fig:jet}(b).

We can measure both the jet diameter $d_{\rm jet}$ and the drop diameter $d_{\rm drop}$ 
in the jetting regime. From these two diameters, we obtain $\lambda_f$, 
the wavelength of the fastest growing mode of the Rayleigh-Plateau instability, 
through the relation 
$\frac{\pi}{4} d_{\rm jet}^2 \lambda_f=\frac{\pi}{6}d_{\rm drop}^3$ 
for drop volume. 
Using the simulation results shown in figure \ref{fig:jet}(a), 
we obtain $d_{\rm drop}\approx 2 d_{\rm jet}$ and hence $\lambda\approx 5.3 d_{\rm jet}$, 
which is in the physically reasonable range. 
It is noted that for dripping-to-jetting transitions dominated by outer flows, 
$d_{\rm drop}\approx 2 d_{\rm jet}$ has been experimentally observed 
\cite{utada2007dripping,mak2017dripping}.  

\subsection{\label{sec:a_dominant} Transitions dominated by inner flows }
In this subsection, we investigate the second class of dripping-to-jetting transitions driven by strong inner flows.
For this purpose, the value of $\bar{v}_{\rm out}$ is fixed to be small, and the value of $\bar{v}_{\rm in}$ is increased to induce the transition. 

For $\bar{R}=0.1$, we have $\bar{v}_{\rm out}=0.29\bar b$ and 
$\mathcal{C}_{\rm out}=0.029$ for the typical value $0.1$ used for $\bar b$. 

\begin{figure}[!htbp]
 \centering
  \subfigure{ \includegraphics[scale=.52]{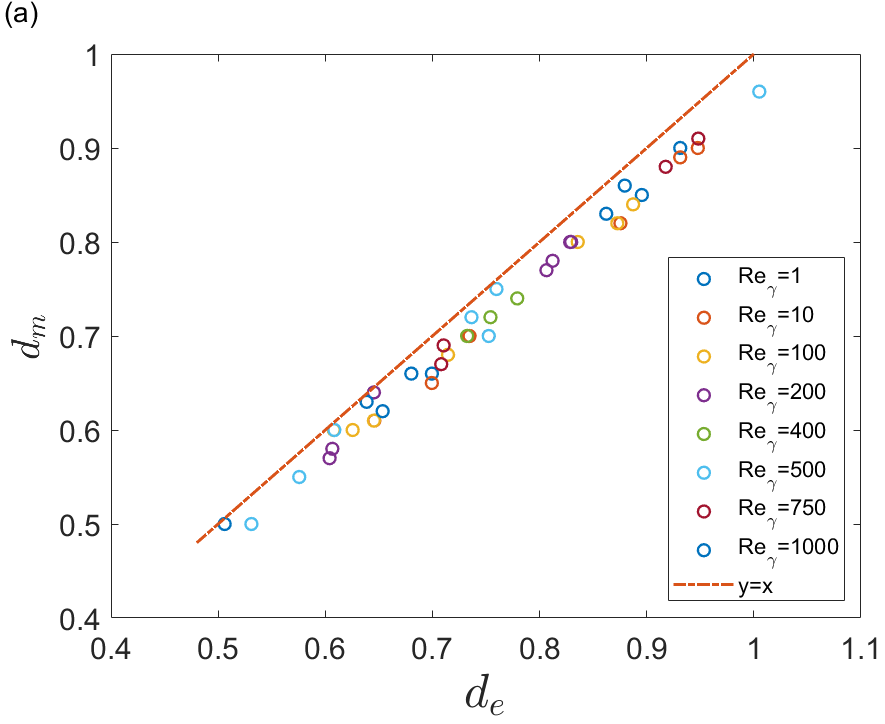}}
  \subfigure{ \includegraphics[scale=.52]{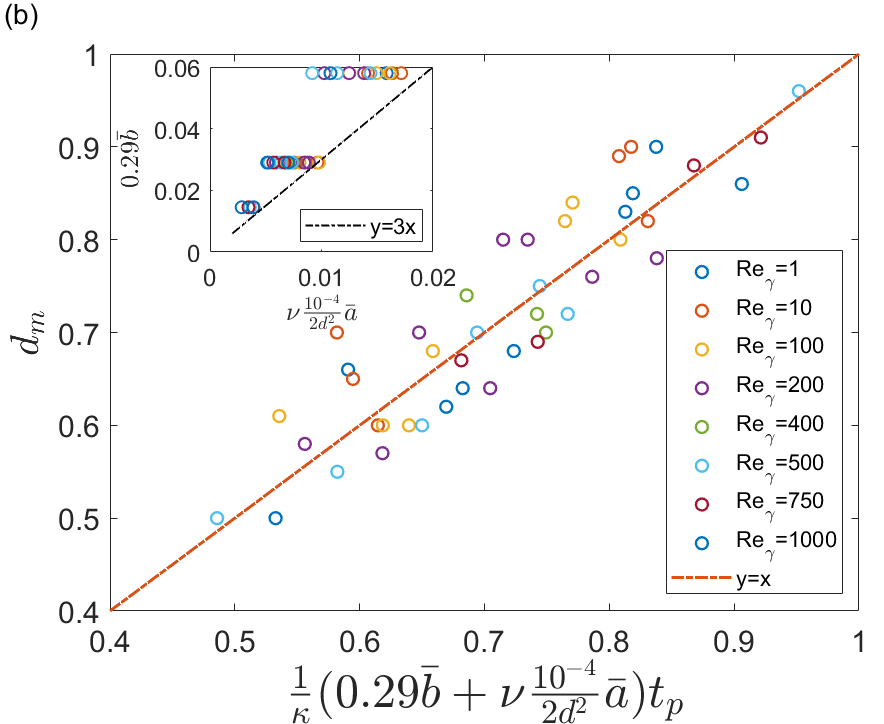}}
 \caption{ (a) A comparison between the diameter $d_e$ expected from mass conservation and
 the diameter $d_m$ measured in our simulations.
 Note that $d_m$ is always slightly smaller than $d_e$ due to the bulk diffusion.  %%%%%%%%%
  (b) The relation between the drop diameter $d_m$ and the time period $t_p$ of drop generation.
  Here the parameters $\bar a$ and $\bar b$, which control $\bar{v}_{\rm in}$ and $\bar{v}_{\rm out}$, are also involved
  according to equations (\ref{eq:tp_0}) and (\ref{eq:tp_1}), with $\kappa=3.1$ and $\nu=3$.
  The inset shows that the contribution of $0.29 \bar b$ is much larger than
  that of $\nu \frac{10^{-4}}{2 d^2}\bar a$ in equation (\ref{eq:tp_1}),
  indicating that the growing drop is mainly advected by the outer flow.
  The data are obtained by using $\varepsilon=0.01$, $B=0.0002$, $\bar{R}=0.1$, $\bar{H}=4$,
  and different combinations of $Re_{\gamma}$, $\bar a$ and $\bar b$.
 }\label{fig:Dropsize}
 \end{figure}

We start from the drop size in the dripping regime. When $\bar{v}_{\rm in}$ is not large enough, the system is
in a dripping state in which drops of the same size are periodically generated at the same pinch-off position.
From the periodic dynamics and mass conservation, we obtain $\frac{4}{3}\pi (\frac{d_e}{2})^3=\pi\bar{R}^2 \bar{v}_{\rm in} t_p$,
where $\bar{R}$ is the radius of the orifice, $\bar{v}_{\rm in}$ is the average velocity of the inner fluid,
$t_p$ is the time period of the periodic generation of drops,
and $d_e$ is the diameter of the drops expected from mass conservation.
Figure \ref{fig:Dropsize}(a) shows a comparison between the expected diameter $d_{ e}$ and the diameter $d_{m}$
which is measured in our numerical simulations. It is noted that in each simulation,
$d_m$ is slightly smaller than $d_e$ expected from mass conservation.
This is attributed to the bulk diffusion which continuously reduce the size of drops.

To understand how the drop size is controlled by the inner and outer flows, we show that
the time period $t_p$ can be related to the drop diameter $d$ as follows:
 \begin{equation}\label{eq:tp_0}
 t_p \approx \frac{\kappa d}{\bar{v}_{\rm out}+\nu \frac{\bar{R}^2}{d^2} \bar{v}_{\rm in}}
 \end{equation}
where $\mu$ and $\nu$ are two adjustable parameters of the order of magnitude of $1$,
and $\bar{v}_{\rm in}$ and $\bar{v}_{\rm out}$ have been defined in \eqref{v_in_out}.
For $\bar{R}=0.1$, we have
 \begin{equation}\label{eq:tp_1}
 t_p \approx \frac{\kappa d}{(0.29 \bar b + \nu \frac{10^{-4}}{2 d^2}\bar a)},
 \end{equation}
which has been numerically verified by figure \ref{fig:Dropsize}(b) in which the measured diameter $d_m$
is used for the drop diameter $d$.
Physically, equation (\ref{eq:tp_0}) describes the advection of a growing drop, 
with the advected distance being $\sim d$ and the velocity being 
$\sim \bar{v}_{\rm out}+\nu \frac{\bar{R}^2}{d^2} \bar{v}_{\rm in}$, 
in which the contribution of $\bar{v}_{\rm in}$ is rescaled by 
a factor $\sim \frac{\bar{R}^2}{d^2}$. 
Equation (\ref{eq:tp_1}) is then obtained by using equation (\ref{v_in_out}) 
to express $\bar{v}_{\rm in}$ and $\bar{v}_{\rm out}$ for $\bar{R}=0.1$. 
From our simulation results, the data points in figure \ref{fig:Dropsize}(b) 
are produced by using optimal values for the adjustable parameters $\kappa$ and $\nu$ 
to best fit the solid line representing equation (\ref{eq:tp_1}). 
Furthermore, it is seen from the inset to figure \ref{fig:Dropsize}(b) that
the contribution of $\bar{v}_{\rm out}$ is much larger than that of $\nu \frac{\bar{R}^2}{d^2} \bar{v}_{\rm in}$
in equation (\ref{eq:tp_0}), i.e.,
the contribution of $0.29 \bar{b}$ is much larger than that of $\nu \frac{10^{-4}}{2 d^2}\bar a$
in equation (\ref{eq:tp_1}) for $\bar{R}=0.1$.
This means that for the advection of a growing drop, the distance is typically $\sim d$, and the velocity is
predominantly $\sim \bar{v}_{\rm out}$. It follows that the time period $t_p$ of drop generation is
$\sim \frac{d}{\bar{v}_{\rm out}}$. Combining $t_p\sim \frac{d}{\bar{v}_{\rm out}}$ and
$\frac{4}{3}\pi (\frac{d}{2})^3=\pi\bar{R}^2 \bar{v}_{\rm in} t_p$ from mass conservation, we have
$d\sim \bar{R}\sqrt{\frac{\bar{v}_{\rm in}}{\bar{v}_{\rm out}}}$, 
 which has been experimentally verified \cite{utada2007dripping}.

%%%%%%%%%

\begin{figure}[!htbp]
% \captionsetup[subfigure]{labelformat=empty}
 \centering
%\subfigure[Slope VS $\xi $ ]{ \includegraphics[scale=.3]{SlopeVSxi.png}} HFK: Figure 21 (a) & (b)
 % \subfigure[]{ \includegraphics[scale=.32]{Dripping_a24.png}}
 % \subfigure[]{ \includegraphics[scale=.32]{Jetting_a25.png}}\\
  \subfigure{ \includegraphics[scale=.33]{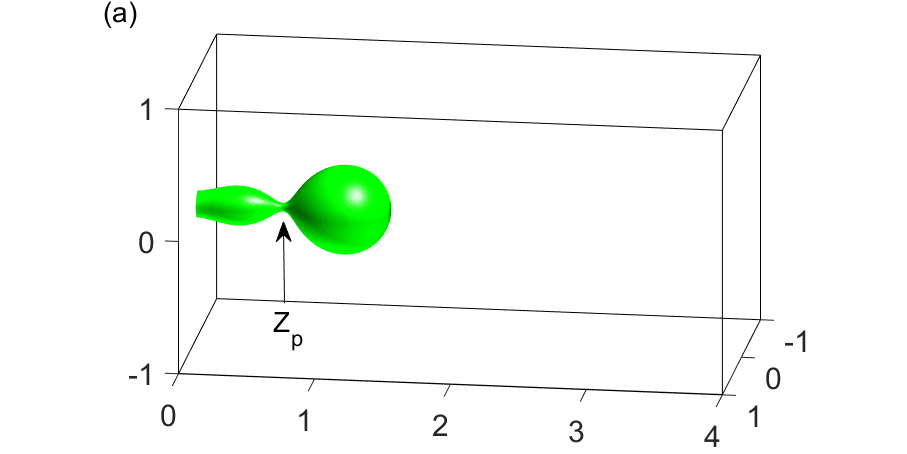}}
  \subfigure{ \includegraphics[scale=.33]{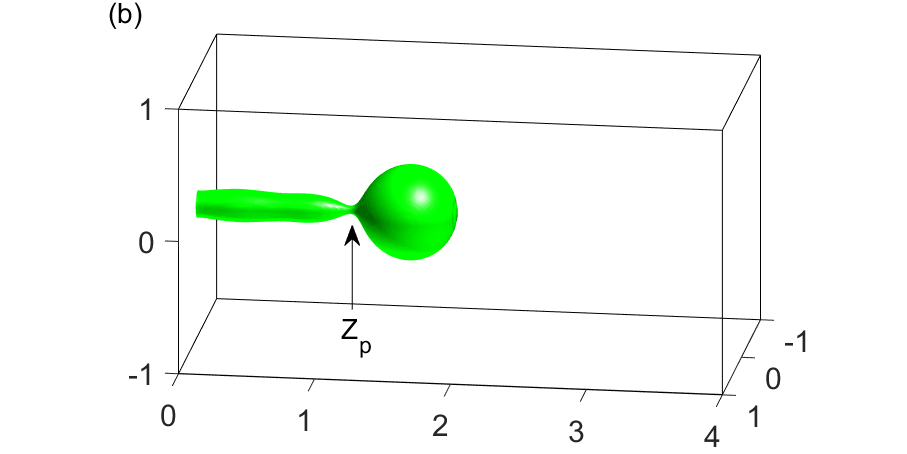}}\\
  \subfigure{ \includegraphics[scale=.41]{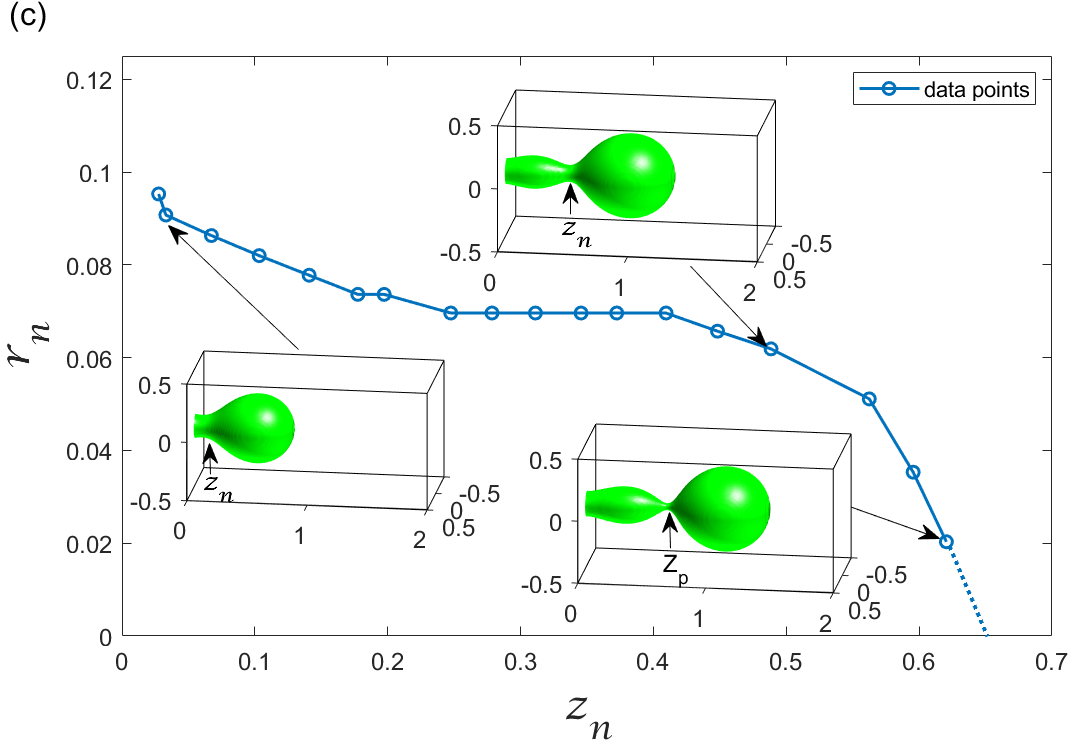}}
  \subfigure{ \includegraphics[scale=.41]{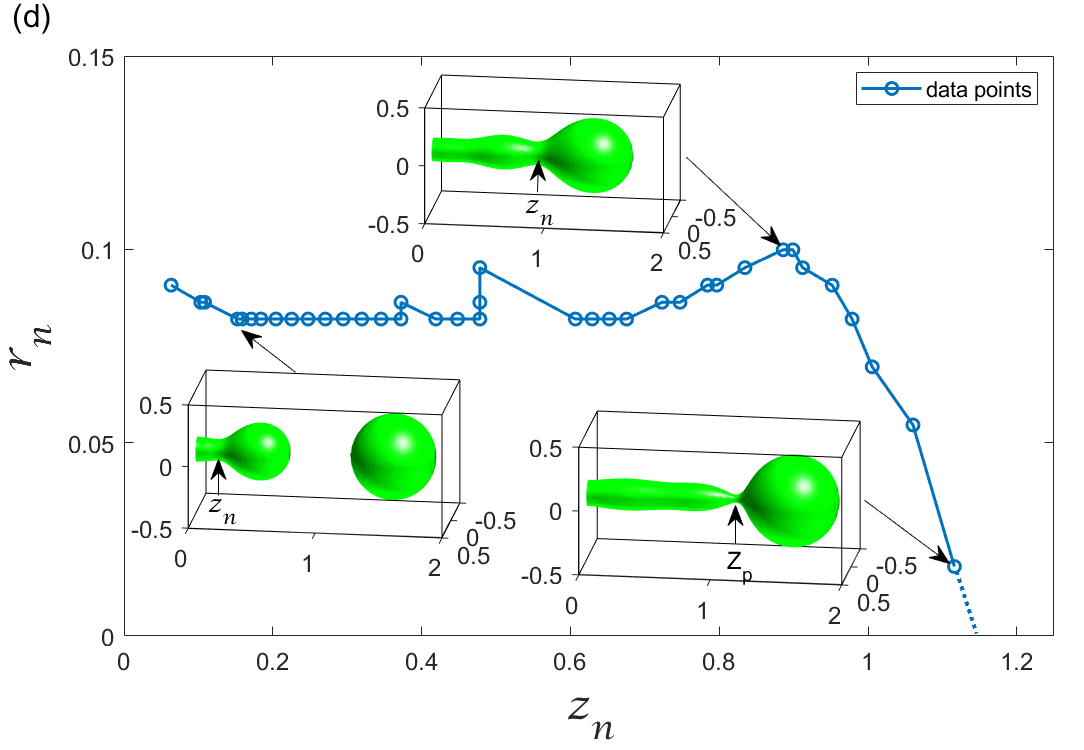}}
 \caption{(a)-(b)
 Two different pinch-off positions for two different inner flow rates,
 with $\bar a=24$ for a dripping state in (a) and $\bar a=25$ for a jetting state in (b).
 (c)-(d) Variation of the neck radius $r_n$ with the neck position $z_n$ as time goes on.
 Here the third inset to (c) corresponds to (a) for dripping, and the third inset to (d) corresponds to (b) for jetting.
 It is noted in (d) that before $r_n$ eventually decreases to $0$, it exhibits a transient increase that
 leads to a visible jump in the value of $Z_p$.
 The data are obtained by using $ \varepsilon=0.01$, $Re_{\gamma}=500$, $B=0.0002$, $\bar{R}=0.1$, $\bar{H}=4$, and $\bar b=0.1$.
 }\label{fig:P}
 \end{figure}

Now we focus on the dripping-to-jetting transitions dominated by inner flows. Same as done in the previous subsection,
we use $Z_p$ to denote the distance between the pinch-off position and the boundary of $\bar{z}=0$ (the orifice).
For a slow outer flow fixed at $\bar b=0.1$, $Z_p$ is expected to increase with the increasing inner flow rate,
i.e., the increasing $\bar a$.
Figure \ref{fig:P}(a) shows the pinch-off position in a dripping state for $\bar a=24$ just before the transition,
and figure \ref{fig:P}(b) shows the pinch-off position in a jetting state for $\bar a=25$ just after the transition.
It is clearly observed that there is a sharp increase of $Z_p$ from figure \ref{fig:P}(a) to \ref{fig:P}(b),
indicating the occurrence of a dripping-to-jetting transition.
Here the value of $Re_\gamma$ is $500$, and
we have $\mathcal{W}_{\rm in}=0.72$ and $\mathcal{C}_{\rm in}=0.12$ for $\bar a=24$,
and $\mathcal{W}_{\rm in}=0.781$ and $\mathcal{C}_{\rm in}=0.125$ for $\bar a=25$.
It is noted that the value of $Re_\gamma$ used here is large enough to let $\mathcal{W}_{\rm in}$ be in control,
with $\mathcal{C}_{\rm in}$ being less important.
It is also noted that the critical magnitude of $\mathcal{W}_{\rm in}$ is in agreement with
the experimental results in \cite{utada2007dripping}.
%%%%%%%%%
According to the state diagram reported in \cite{utada2007dripping}, for the dripping-to-jetting transitions dominated by inner flows, the critical values of $\mathcal{W}_{\rm in}$ are typically distributed around $1$.

To understand the underlying mechanism of the dripping-to-jetting transitions
dominated by inner flows, we use figures \ref{fig:P}(c) and \ref{fig:P}(d) to
show the variation of the neck radius $r_n$ with the neck position $z_n$ as time goes on.
Note that as the neck radius approaches $0$, i.e., $r_n \to 0$,
pinch-off occurs with the neck position approaching the pinch-off position, i.e., $z_n \to Z_p$.
In the dripping regime, it is observed that $r_n$ decreases monotonically to $0$ (as shown in figure \ref{fig:P}(c)),
while in the jetting regime, $r_n$ exhibits a transient increase before it eventually decreases to $0$
(as shown in figure \ref{fig:P}(d)). It is this transient increase of $r_n$ that leads to a visible jump
in the value of $Z_p$ that marks the transition from dripping to jetting.

\begin{figure}[!htbp]
 \centering
{ \includegraphics[scale=.35]{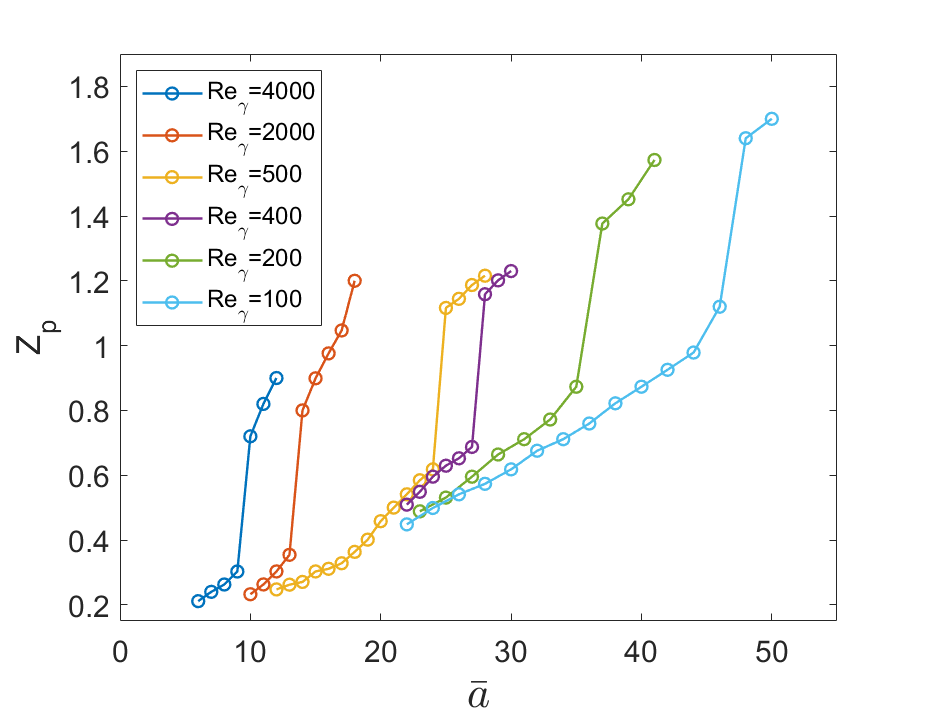}}
 \caption{ Variation of the pinch-off position $Z_p$ with the parameter $\bar{a}$ which controls the inner flow rate.
 For each value of $Re_\gamma$, a transition is noted around a critical value of $\bar{a}$.
 The data are obtained by using $\varepsilon =0.01$, $B=0.0002$, $\bar{R}=0.1$, $\bar{H}=4$, and $\bar b=0.1$.}
 \label{fig:Jump}
 \end{figure}

Figure \ref{fig:Jump} shows the variation of the pinch-off position $Z_p$ with the parameter $a$
which controls the inner flow rate. For each value of $Re_\gamma$, a transition is noted around
a critical value of $a$. Furthermore, this critical value of $a$ increases with the decreasing $Re_\gamma$.

The jump in the pinch-off position is a clear indicator that can be used to locate the dripping-to-jetting transition.
In the following, we focus on the critical velocity of the inner flow that is needed to induce the transition,
with the interfacial tension $\gamma$ being varied for nearly two orders of magnitude. %%%%%%%%%
The dripping-to-jetting transitions in systems of high interfacial tension have been extensively studied
\cite{utada2007dripping}. In particular, when the transition is dominated by the inner flow
(with the outer flow rate measured by $\mathcal{C}_{\rm out}$ being negligible),
the inertial force due to the inner flow, measured by $\mathcal{W}_{\rm in}$, plays a dominant role
in systems of high interfacial tension.
However, when the interfacial tension is continuously lowered,
the viscous force due to the inner flow, measured by $\mathcal{C}_{\rm in}$,
becomes more and more important in driving the transition.
This trend has been reported experimentally \cite{mak2017dripping},
and a theoretical understanding can be described as follows.
The Weber number of the inner flow is given by $\mathcal{W}_{\rm in}=\bar{v}_{\rm in}^2 Re_{\gamma}\bar{R}$,
where the interfacial tension $\gamma$ is involved in the Reynolds number $Re_{\gamma}$ defined by
$Re_{\gamma}=\frac{\rho u L}{\eta}$, with $u=\frac{\gamma}{\eta}$ being the velocity unit.
Let's suppose that the transition occurs at $\mathcal{W}_{\rm in} \approx 1$,
with the interfacial tension force being balanced by the inertial force due to the inner flow.
If $Re_{\gamma}$ is made sufficiently small by a sufficiently low interfacial tension,
then the value of $\bar{v}_{\rm in}$ corresponding to $\mathcal{W}_{\rm in}=\bar{v}_{\rm in}^2 Re_{\gamma}\bar{R}\approx 1$
can be made large enough to be comparable to $\mathcal{W}_{\rm in}$.
Note that the capillary number of the inner flow is given by $\mathcal{C}_{\rm in}=\bar{v}_{\rm in}$.
With $\mathcal{C}_{\rm in}$ being comparable to $\mathcal{W}_{\rm in}\approx 1$ for sufficiently low interfacial tension,
it is deduced that the viscous force due to the inner flow is no longer negligible compared to
the inertial force in driving the transition in systems of low interfacial tension.

\begin{figure}[!htbp]
 \centering
 \subfigure{ \includegraphics[scale=.5]{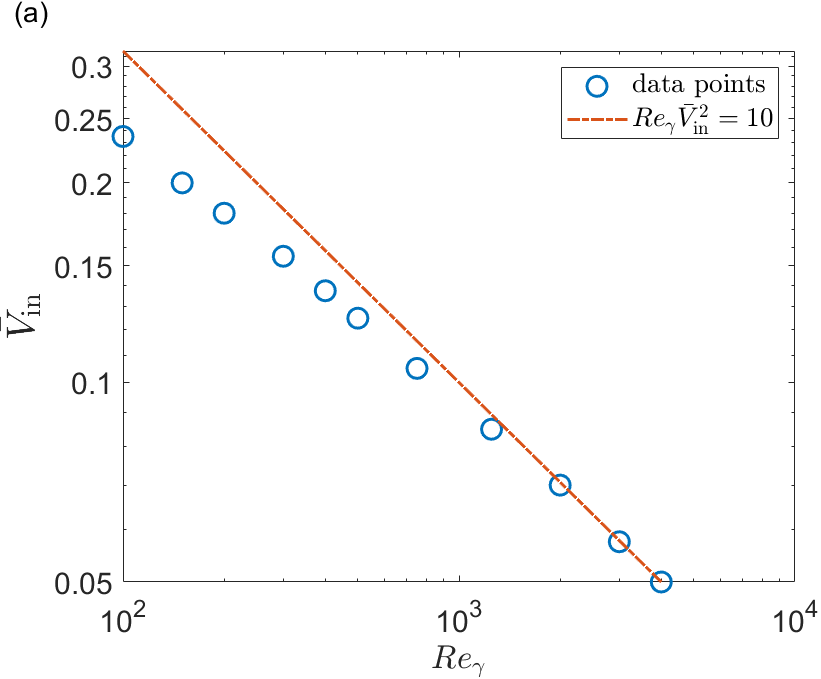}}
 \subfigure{ \includegraphics[scale=.5]{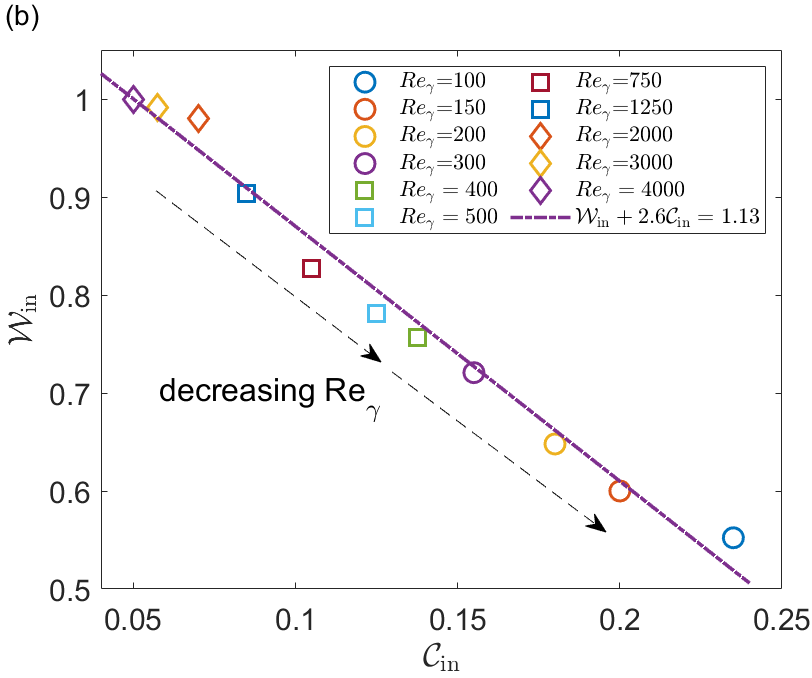}}
 \caption{(a) Log-log plot for the critical velocity of the inner flow $\bar{V}_{\rm in}$ versus
 the Reynolds number $Re_{\gamma}$.
 For large $Re_{\gamma}$, $\bar{V}_{\rm in}^2 Re_{\gamma}={\rm const.}$,
 while for small $Re_{\gamma}$, the critical $\bar{V}_{\rm in}$ becomes smaller than that predicted by
 a constant $\bar{V}_{\rm in}^2 Re_{\gamma}$.
 (b)
 The Weber number of the inner flow $\mathcal{W}_{\rm in}$ and
 the capillary number of the inner flow $\mathcal{C}_{\rm in}$ at the transition
 satisfy $\mathcal{W}_{\rm in}+2.6\mathcal{C}_{\rm in}=1.13$ approximately.
 The data are obtained by using $\varepsilon=0.01$, $B=0.0002$, $\bar{R}=0.1$, $\bar{H}=4$, and
 $\bar b=0.1$, with $Re_{\gamma}$ being varied between $100$ and $4000$.
 A thin dotted line with arrow is used to indicate the direction of change of $Re_{\gamma}$. 
 Note that $\bar b=0.1$ used here gives $\mathcal{C}_{\rm out}=0.029$,
 which is much smaller than the typical values of $\mathcal{W}_{\rm in}$ and $\mathcal{C}_{\rm in}$
 at the transition.}
 \label{fig:Transition}
 \end{figure}

Let $\bar{V}_{\rm in}$ denote the critical velocity of the inner flow. In systems of high interfacial tension,
the inertial force due to the inner flow is dominant, and hence the transition occurs at
$\bar{V}_{\rm in}^2 Re_{\gamma}\bar{R}\approx 1$ for the critical Weber number $\mathcal{W}_{\rm in}\approx 1$.
As a result, $\bar{V}_{\rm in}^2 Re_{\gamma}={\rm const.}$ is expected for large $Re_{\gamma}$.
This is indeed observed in figure \ref{fig:Transition}(a).
When $Re_{\gamma}$ is no longer large enough, deviation from $\bar{V}_{\rm in}^2 Re_{\gamma}={\rm const.}$
does show up. From figure \ref{fig:Transition}(a), it is seen that toward the low end of the range of $Re_{\gamma}$,
the critical $\bar{V}_{\rm in}$ is actually below that predicted by $\bar{V}_{\rm in}^2 Re_{\gamma}={\rm const.}$,
which only considers the inertial force due to the inner flow.
As explained above, when the interfacial tension is low and hence $Re_{\gamma}$ is small, the value of $\bar{V}_{\rm in}$ predicted by $\bar{V}_{\rm in}^2 Re_{\gamma}={\rm const.}$ is large. This means a large viscous force due to the inner flow. As a result, the viscous force and inertial force due to the inner flow are added up to jointly balance the interfacial tension force. Consequently, the critical $\bar{V}_{\rm in}$ becomes smaller than that predicted by $\bar{V}_{\rm in}^2 Re_{\gamma}={\rm const.}$, which only considers the inertial force due to the inner flow.

%%%%%%%%%
For $Re_{\gamma}$ being varied between $100$ and $4000$, numerical simulations have been carried out to determine
the critical velocity of the inner flow $\bar{V}_{\rm in}$ at which the transition occurs.
The data obtained for $\bar{V}_{\rm in}$ are used to produce a formula that describes the contributions of
the Weber number of the inner flow $\mathcal{W}_{\rm in}$ and
the capillary number of the inner flow $\mathcal{C}_{\rm in}$ at the transition.
Figure \ref{fig:Transition}(b) shows that $\mathcal{W}_{\rm in}$ and $\mathcal{C}_{\rm in}$ at the transition
satisfy a linear relation given by $\mathcal{W}_{\rm in}+2.6\mathcal{C}_{\rm in}=1.13$ approximately.
It is worth emphasizing that this equation holds for
the interfacial tension $\gamma$ being varied for nearly two orders of magnitude.
%%% Question for Huang FK: In figure \ref{fig:Transition}(b), data toward the lower right end are obtained for
%%% small $Re_\gamma$, and data toward the upper left end are obtained for large $Re_\gamma$. Is that right?
%%% If yes, then it should be pointed out explicitly.
Note that from the upper left to the lower right, the value of $Re_\gamma$ decreases and consequently
the relative importance of $\mathcal{C}_{\rm in}$ increases. Therefore, it is numerically verified that
the viscous force due to the inner flow plays a quantitatively important role in driving
the dripping-to-jetting transitions in systems of low interfacial tension.
%%%%%%%%%

%\begin{figure}[!htbp]
% \centering
%{ \includegraphics[scale=.4]{Transition.png}}
% \caption{ The relation between the Weber number and the capillary number of the inner flow when transition happens, showing that  the Weber number and the capillary number of the inner flow approximately satisfy $\mathcal{W}_{in}+2.6\mathcal{C}a_{out}=1.13$.  The data are obtained by choosing
% $\bar{R}=0.1$, $\bar{H}=4$, $B=0.0002$ and varying $b=0.05, 0.1, 0.2$. The range for $Re_{\gamma}$ is from $100 \sim 4000$.  }.
% \label{fig:Transition}
 %\end{figure}

\subsection{\label{sec:B_effect} Effect of bulk diffusion }

\begin{figure}[!htbp]
 \centering
{ \includegraphics[scale=.38]{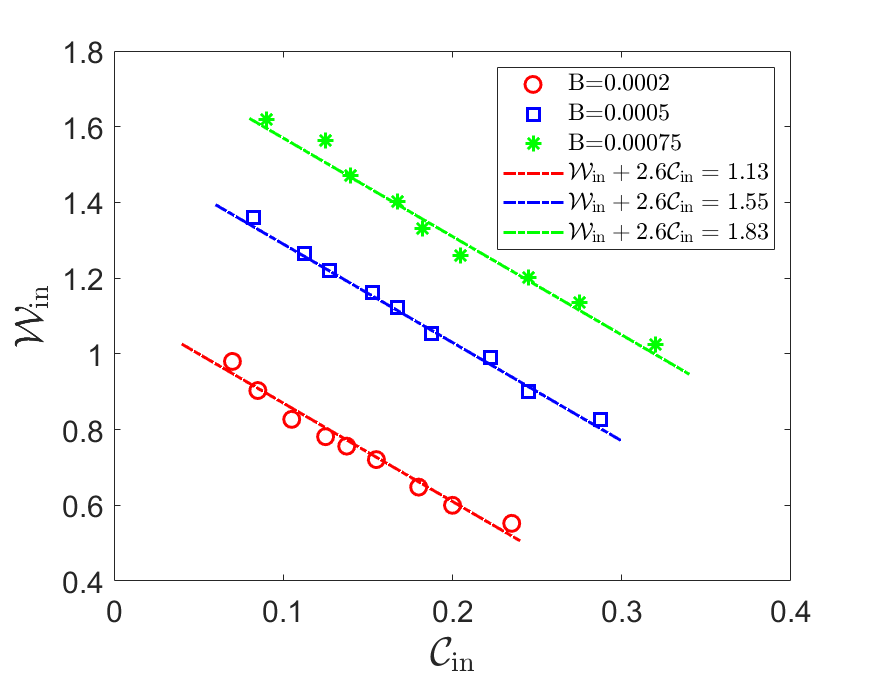}}
 \caption{ The Weber number of the inner flow $\mathcal{W}_{\rm in}$ and
 the capillary number of the inner flow $\mathcal{C}_{\rm in}$ at the transition
 always satisfy a linear relation for transitions dominated by inner flows.
 In addition, the three fitting lines for three different values of $B$ are parallel.
 The data are obtained by using $ \varepsilon=0.01$, $\bar{R}=0.1$, $\bar{H}=4$, $\bar b=0.1$, with
 $Re_{\gamma}$ being varied between $100$ and $2000$, and $B=0.0002$, $0.0005$, and $0.00075$.
 The value of $Re_\gamma$ decreases from the upper left to the lower right along each line.
 Note that $\bar b=0.1$ used here gives $\mathcal{C}_{\rm out}=0.029$,
 which is much smaller than the typical values of $\mathcal{W}_{\rm in}$ and $\mathcal{C}_{\rm in}$
 at the transition.}
 \label{fig:TransitionB}
 \end{figure}

In this subsection, we investigate the effect of bulk diffusion
on the condition for the occurrence of transition.
%%%%%%%%%
Physically, bulk diffusion is a dissipative process that can lower the interfacial energy and lead to
the breakup of a liquid thread \cite{Xu2019,huang2022diffuse}.
Therefore, adding bulk diffusion to the system will facilitate the pinch-off dynamics and
hence hinder the development of jetting state.
As a result, a larger critical velocity $\bar{V}_{\rm in}$ is needed to induce the dripping-to-jetting transition.
In an earlier work \cite{huang2022diffuse}, we demonstrate that the effect of bulk diffusion can be enhanced by
increasing the characteristic length scale $l_c$,
which enters into the dimensionless system through the parameter $B=\frac{2 l_c^2}{L^2}$.
Figure \ref{fig:TransitionB} shows that at different levels of bulk diffusion controlled by $B$,
$\mathcal{W}_{\rm in}$ and $\mathcal{C}_{\rm in}$ at the transition always satisfy a linear relation
for transitions dominated by inner flow. Two important observations are made from Figure \ref{fig:TransitionB}.
(i) Stronger diffusion indeed necessitates a larger critical velocity $\bar{V}_{\rm in}$ to induce the transition.
(ii) The three fitting lines are parallel, showing that the relative contributions of
the inertial force and viscous force due to the inner flow remain the same
regardless of the variation of bulk diffusion.

\section{\label{sec:conclusion} Concluding remarks}

The CHNS model has been solved in a cylindrical domain with axisymmetry to investigate the dripping-to-jetting transitions
in coaxial flows of two immiscible fluids.
Numerous numerical examples are presented to demonstrate that the distance between the orifice and pinch-off position
increases when either the outer or the inner flow rate is enhanced.
It is observed that there is an apparent jump in this distance when the outer or the inner flow rate reaches the critical value
for the dripping-to-jetting transition to occur.
The critical flow rates numerically obtained for both the outer and inner flows are consistent with
the corresponding experimental results in order of magnitude.
For transitions dominated by outer flows, a thin and long jet is generated when jetting occurs,
and our numerical results for the jet radius are validated by its dependence on the outer flow rate
according to the mass conservation.
For transitions dominated by inner flows, the interfacial tension is varied for nearly two orders of magnitude,
and a quantitative relation is established between the contributions of the inertial and viscous forces due to the inner flow
at the transition point.
Finally, the degree of bulk diffusion is varied to show its quantitative effect
on the critical flow rate at the transition point.

To the best of our knowledge, there has been no prior work that employs a phase-field model to investigate
the dripping-to-jetting transitions in three dimensions, with a focus on the effects of low interfacial tension and bulk diffusion.
In the present work, we have considered the simplest situation in which the two fluids have equal density, equal viscosity and
equal diffusion coefficient. Actually, these restrictions can be lifted
in both experiments \cite{utada2007dripping,mak2017dripping,Xu2019} and numerical simulations \cite{dong2012}.
Although the dripping-to-jetting transitions for high interfacial tension have been extensively studied in the past two decades,
low interfacial tension and bulk diffusion may inject new ingredients into this classical problem.
In this regard, quantitative effects of density ratio, viscosity ratio and diffusivity ratio
largely remain to be explored in both experiments and numerical simulations.

\begin{acknowledgments}
The work of F. Huang and W. Bao was supported by the Ministry of Education of Singapore under its AcRF Tier 2 funding MOE-T2EP20122-0002(A-8000962-00-00),
and the work of T. Qian was supported by the Hong Kong RGC grants CRF No. C1006-20WF and GRF No. 16306121. T. Qian was also supported by the Key Project of the National Natural Science Foundation of China (No. 12131010). Part of this work was done when the first two authors were visiting the Institute for Mathematics Sciences at the National University of Singapore in February 2023.
\end{acknowledgments}

\section*{Data Availability Statement}
The data that support the findings of this study are available from the corresponding author
upon reasonable request.
\section*{References}
\nocite{*}
\bibliography{drip.bib}% Produces the bibliography via BibTeX.

%merlin.mbs aipnum4-1.bst 2010-07-25 4.21a (PWD, AO, DPC) hacked
%Control: key (0)
%Control: author (8) initials jnrlst
%Control: editor formatted (1) identically to author
%Control: production of article title (0) allowed
%Control: page (1) range
%Control: year (1) truncated
%Control: production of eprint (0) enabled
\begin{thebibliography}{33}%
\makeatletter
\providecommand \@ifxundefined [1]{%
 \@ifx{#1\undefined}
}%
\providecommand \@ifnum [1]{%
 \ifnum #1\expandafter \@firstoftwo
 \else \expandafter \@secondoftwo
 \fi
}%
\providecommand \@ifx [1]{%
 \ifx #1\expandafter \@firstoftwo
 \else \expandafter \@secondoftwo
 \fi
}%
\providecommand \natexlab [1]{#1}%
\providecommand \enquote  [1]{``#1''}%
\providecommand \bibnamefont  [1]{#1}%
\providecommand \bibfnamefont [1]{#1}%
\providecommand \citenamefont [1]{#1}%
\providecommand \href@noop [0]{\@secondoftwo}%
\providecommand \href [0]{\begingroup \@sanitize@url \@href}%
\providecommand \@href[1]{\@@startlink{#1}\@@href}%
\providecommand \@@href[1]{\endgroup#1\@@endlink}%
\providecommand \@sanitize@url [0]{\catcode `\\12\catcode `\$12\catcode
  `\&12\catcode `\#12\catcode `\^12\catcode `\_12\catcode `\%12\relax}%
\providecommand \@@startlink[1]{}%
\providecommand \@@endlink[0]{}%
\providecommand \url  [0]{\begingroup\@sanitize@url \@url }%
\providecommand \@url [1]{\endgroup\@href {#1}{\urlprefix }}%
\providecommand \urlprefix  [0]{URL }%
\providecommand \Eprint [0]{\href }%
\providecommand \doibase [0]{http://dx.doi.org/}%
\providecommand \selectlanguage [0]{\@gobble}%
\providecommand \bibinfo  [0]{\@secondoftwo}%
\providecommand \bibfield  [0]{\@secondoftwo}%
\providecommand \translation [1]{[#1]}%
\providecommand \BibitemOpen [0]{}%
\providecommand \bibitemStop [0]{}%
\providecommand \bibitemNoStop [0]{.\EOS\space}%
\providecommand \EOS [0]{\spacefactor3000\relax}%
\providecommand \BibitemShut  [1]{\csname bibitem#1\endcsname}%
\let\auto@bib@innerbib\@empty
%</preamble>
\bibitem [{\citenamefont {Whitesides}(2006)}]{whitesides2006origins}%
  \BibitemOpen
  \bibfield  {author} {\bibinfo {author} {\bibfnamefont {G.~M.}\ \bibnamefont
  {Whitesides}},\ }\bibfield  {title} {\enquote {\bibinfo {title} {The origins
  and the future of microfluidics},}\ }\href@noop {} {\bibfield  {journal}
  {\bibinfo  {journal} {Nature}\ }\textbf {\bibinfo {volume} {442}},\ \bibinfo
  {pages} {368--373} (\bibinfo {year} {2006})}\BibitemShut {NoStop}%
\bibitem [{\citenamefont {Stone}, \citenamefont {Stroock},\ and\ \citenamefont
  {Ajdari}(2004)}]{stone2004engineering}%
  \BibitemOpen
  \bibfield  {author} {\bibinfo {author} {\bibfnamefont {H.~A.}\ \bibnamefont
  {Stone}}, \bibinfo {author} {\bibfnamefont {A.~D.}\ \bibnamefont {Stroock}},
  \ and\ \bibinfo {author} {\bibfnamefont {A.}~\bibnamefont {Ajdari}},\
  }\bibfield  {title} {\enquote {\bibinfo {title} {Engineering flows in small
  devices: microfluidics toward a lab-on-a-chip},}\ }\href@noop {} {\bibfield
  {journal} {\bibinfo  {journal} {Annu. Rev. Fluid Mech.}\ }\textbf {\bibinfo
  {volume} {36}},\ \bibinfo {pages} {381--411} (\bibinfo {year}
  {2004})}\BibitemShut {NoStop}%
\bibitem [{\citenamefont {Marre}\ \emph {et~al.}(2009)\citenamefont {Marre},
  \citenamefont {Aymonier}, \citenamefont {Subra},\ and\ \citenamefont
  {Mignard}}]{marre2009dripping}%
  \BibitemOpen
  \bibfield  {author} {\bibinfo {author} {\bibfnamefont {S.}~\bibnamefont
  {Marre}}, \bibinfo {author} {\bibfnamefont {C.}~\bibnamefont {Aymonier}},
  \bibinfo {author} {\bibfnamefont {P.}~\bibnamefont {Subra}}, \ and\ \bibinfo
  {author} {\bibfnamefont {E.}~\bibnamefont {Mignard}},\ }\bibfield  {title}
  {\enquote {\bibinfo {title} {Dripping to jetting transitions observed from
  supercritical fluid in liquid microcoflows},}\ }\href@noop {} {\bibfield
  {journal} {\bibinfo  {journal} {Appl. Phys. Lett.}\ }\textbf {\bibinfo
  {volume} {95}},\ \bibinfo {pages} {134105} (\bibinfo {year}
  {2009})}\BibitemShut {NoStop}%
\bibitem [{\citenamefont {Kaufman}\ \emph {et~al.}(2012)\citenamefont
  {Kaufman}, \citenamefont {Tao}, \citenamefont {Shabahang}, \citenamefont
  {Banaei}, \citenamefont {Deng}, \citenamefont {Liang}, \citenamefont
  {Johnson}, \citenamefont {Fink},\ and\ \citenamefont
  {Abouraddy}}]{kaufman2012structured}%
  \BibitemOpen
  \bibfield  {author} {\bibinfo {author} {\bibfnamefont {J.~J.}\ \bibnamefont
  {Kaufman}}, \bibinfo {author} {\bibfnamefont {G.}~\bibnamefont {Tao}},
  \bibinfo {author} {\bibfnamefont {S.}~\bibnamefont {Shabahang}}, \bibinfo
  {author} {\bibfnamefont {E.-H.}\ \bibnamefont {Banaei}}, \bibinfo {author}
  {\bibfnamefont {D.~S.}\ \bibnamefont {Deng}}, \bibinfo {author}
  {\bibfnamefont {X.}~\bibnamefont {Liang}}, \bibinfo {author} {\bibfnamefont
  {S.~G.}\ \bibnamefont {Johnson}}, \bibinfo {author} {\bibfnamefont
  {Y.}~\bibnamefont {Fink}}, \ and\ \bibinfo {author} {\bibfnamefont {A.~F.}\
  \bibnamefont {Abouraddy}},\ }\bibfield  {title} {\enquote {\bibinfo {title}
  {Structured spheres generated by an in-fibre fluid instability},}\
  }\href@noop {} {\bibfield  {journal} {\bibinfo  {journal} {Nature}\ }\textbf
  {\bibinfo {volume} {487}},\ \bibinfo {pages} {463--467} (\bibinfo {year}
  {2012})}\BibitemShut {NoStop}%
\bibitem [{\citenamefont {Utada}\ \emph {et~al.}(2007)\citenamefont {Utada},
  \citenamefont {Fernandez-Nieves}, \citenamefont {Stone},\ and\ \citenamefont
  {Weitz}}]{utada2007dripping}%
  \BibitemOpen
  \bibfield  {author} {\bibinfo {author} {\bibfnamefont {A.~S.}\ \bibnamefont
  {Utada}}, \bibinfo {author} {\bibfnamefont {A.}~\bibnamefont
  {Fernandez-Nieves}}, \bibinfo {author} {\bibfnamefont {H.~A.}\ \bibnamefont
  {Stone}}, \ and\ \bibinfo {author} {\bibfnamefont {D.~A.}\ \bibnamefont
  {Weitz}},\ }\bibfield  {title} {\enquote {\bibinfo {title} {Dripping to
  jetting transitions in coflowing liquid streams},}\ }\href@noop {} {\bibfield
   {journal} {\bibinfo  {journal} {Phys. Rev. Lett.}\ }\textbf {\bibinfo
  {volume} {99}},\ \bibinfo {pages} {094502} (\bibinfo {year}
  {2007})}\BibitemShut {NoStop}%
\bibitem [{\citenamefont {Guillot}\ \emph {et~al.}(2007)\citenamefont
  {Guillot}, \citenamefont {Colin}, \citenamefont {Utada},\ and\ \citenamefont
  {Ajdari}}]{guillot2007stability}%
  \BibitemOpen
  \bibfield  {author} {\bibinfo {author} {\bibfnamefont {P.}~\bibnamefont
  {Guillot}}, \bibinfo {author} {\bibfnamefont {A.}~\bibnamefont {Colin}},
  \bibinfo {author} {\bibfnamefont {A.~S.}\ \bibnamefont {Utada}}, \ and\
  \bibinfo {author} {\bibfnamefont {A.}~\bibnamefont {Ajdari}},\ }\bibfield
  {title} {\enquote {\bibinfo {title} {Stability of a jet in confined
  pressure-driven biphasic flows at low {R}eynolds numbers},}\ }\href@noop {}
  {\bibfield  {journal} {\bibinfo  {journal} {Phys. Rev. Lett.}\ }\textbf
  {\bibinfo {volume} {99}},\ \bibinfo {pages} {104502} (\bibinfo {year}
  {2007})}\BibitemShut {NoStop}%
\bibitem [{\citenamefont {Castro-Hernandez}\ \emph {et~al.}(2009)\citenamefont
  {Castro-Hernandez}, \citenamefont {Gundabala}, \citenamefont
  {Fern{\'a}ndez-Nieves},\ and\ \citenamefont {Gordillo}}]{castro2009scaling}%
  \BibitemOpen
  \bibfield  {author} {\bibinfo {author} {\bibfnamefont {E.}~\bibnamefont
  {Castro-Hernandez}}, \bibinfo {author} {\bibfnamefont {V.}~\bibnamefont
  {Gundabala}}, \bibinfo {author} {\bibfnamefont {A.}~\bibnamefont
  {Fern{\'a}ndez-Nieves}}, \ and\ \bibinfo {author} {\bibfnamefont {J.~M.}\
  \bibnamefont {Gordillo}},\ }\bibfield  {title} {\enquote {\bibinfo {title}
  {Scaling the drop size in coflow experiments},}\ }\href@noop {} {\bibfield
  {journal} {\bibinfo  {journal} {New J. Phys.}\ }\textbf {\bibinfo {volume}
  {11}},\ \bibinfo {pages} {075021} (\bibinfo {year} {2009})}\BibitemShut
  {NoStop}%
\bibitem [{\citenamefont {Ga{\~n}{\'a}n-Calvo}(1998)}]{ganan1998generation}%
  \BibitemOpen
  \bibfield  {author} {\bibinfo {author} {\bibfnamefont {A.~M.}\ \bibnamefont
  {Ga{\~n}{\'a}n-Calvo}},\ }\bibfield  {title} {\enquote {\bibinfo {title}
  {Generation of steady liquid microthreads and micron-sized monodisperse
  sprays in gas streams},}\ }\href@noop {} {\bibfield  {journal} {\bibinfo
  {journal} {Phys. Rev. Lett.}\ }\textbf {\bibinfo {volume} {80}},\ \bibinfo
  {pages} {285} (\bibinfo {year} {1998})}\BibitemShut {NoStop}%
\bibitem [{\citenamefont {Cubaud}\ and\ \citenamefont
  {Mason}(2008)}]{cubaud2008capillary}%
  \BibitemOpen
  \bibfield  {author} {\bibinfo {author} {\bibfnamefont {T.}~\bibnamefont
  {Cubaud}}\ and\ \bibinfo {author} {\bibfnamefont {T.~G.}\ \bibnamefont
  {Mason}},\ }\bibfield  {title} {\enquote {\bibinfo {title} {Capillary threads
  and viscous droplets in square microchannels},}\ }\href@noop {} {\bibfield
  {journal} {\bibinfo  {journal} {Phys. Fluids}\ }\textbf {\bibinfo {volume}
  {20}},\ \bibinfo {pages} {053302} (\bibinfo {year} {2008})}\BibitemShut
  {NoStop}%
\bibitem [{\citenamefont {Thorsen}\ \emph {et~al.}(2001)\citenamefont
  {Thorsen}, \citenamefont {Roberts}, \citenamefont {Arnold},\ and\
  \citenamefont {Quake}}]{thorsen2001dynamic}%
  \BibitemOpen
  \bibfield  {author} {\bibinfo {author} {\bibfnamefont {T.}~\bibnamefont
  {Thorsen}}, \bibinfo {author} {\bibfnamefont {R.~W.}\ \bibnamefont
  {Roberts}}, \bibinfo {author} {\bibfnamefont {F.~H.}\ \bibnamefont {Arnold}},
  \ and\ \bibinfo {author} {\bibfnamefont {S.~R.}\ \bibnamefont {Quake}},\
  }\bibfield  {title} {\enquote {\bibinfo {title} {Dynamic pattern formation in
  a vesicle-generating microfluidic device},}\ }\href@noop {} {\bibfield
  {journal} {\bibinfo  {journal} {Phys. Rev. Lett.}\ }\textbf {\bibinfo
  {volume} {86}},\ \bibinfo {pages} {4163} (\bibinfo {year}
  {2001})}\BibitemShut {NoStop}%
\bibitem [{\citenamefont {Abate}\ \emph {et~al.}(2009)\citenamefont {Abate},
  \citenamefont {Poitzsch}, \citenamefont {Hwang}, \citenamefont {Lee},
  \citenamefont {Czerwinska},\ and\ \citenamefont {Weitz}}]{abate2009impact}%
  \BibitemOpen
  \bibfield  {author} {\bibinfo {author} {\bibfnamefont {A.}~\bibnamefont
  {Abate}}, \bibinfo {author} {\bibfnamefont {A.}~\bibnamefont {Poitzsch}},
  \bibinfo {author} {\bibfnamefont {Y.}~\bibnamefont {Hwang}}, \bibinfo
  {author} {\bibfnamefont {J.}~\bibnamefont {Lee}}, \bibinfo {author}
  {\bibfnamefont {J.}~\bibnamefont {Czerwinska}}, \ and\ \bibinfo {author}
  {\bibfnamefont {D.}~\bibnamefont {Weitz}},\ }\bibfield  {title} {\enquote
  {\bibinfo {title} {Impact of inlet channel geometry on microfluidic drop
  formation},}\ }\href@noop {} {\bibfield  {journal} {\bibinfo  {journal}
  {Phys. Rev. E}\ }\textbf {\bibinfo {volume} {80}},\ \bibinfo {pages} {026310}
  (\bibinfo {year} {2009})}\BibitemShut {NoStop}%
\bibitem [{\citenamefont {Nunes}\ \emph {et~al.}(2013)\citenamefont {Nunes},
  \citenamefont {Tsai}, \citenamefont {Wan},\ and\ \citenamefont
  {Stone}}]{nunes2013dripping}%
  \BibitemOpen
  \bibfield  {author} {\bibinfo {author} {\bibfnamefont {J.}~\bibnamefont
  {Nunes}}, \bibinfo {author} {\bibfnamefont {S.}~\bibnamefont {Tsai}},
  \bibinfo {author} {\bibfnamefont {J.}~\bibnamefont {Wan}}, \ and\ \bibinfo
  {author} {\bibfnamefont {H.~A.}\ \bibnamefont {Stone}},\ }\bibfield  {title}
  {\enquote {\bibinfo {title} {Dripping and jetting in microfluidic multiphase
  flows applied to particle and fibre synthesis},}\ }\href@noop {} {\bibfield
  {journal} {\bibinfo  {journal} {J. Phys. D .}\ }\textbf {\bibinfo {volume}
  {46}},\ \bibinfo {pages} {114002} (\bibinfo {year} {2013})}\BibitemShut
  {NoStop}%
\bibitem [{\citenamefont {Mak}, \citenamefont {Chao},\ and\ \citenamefont
  {Shum}(2017)}]{mak2017dripping}%
  \BibitemOpen
  \bibfield  {author} {\bibinfo {author} {\bibfnamefont {S.~Y.}\ \bibnamefont
  {Mak}}, \bibinfo {author} {\bibfnamefont {Y.}~\bibnamefont {Chao}}, \ and\
  \bibinfo {author} {\bibfnamefont {H.~C.}\ \bibnamefont {Shum}},\ }\bibfield
  {title} {\enquote {\bibinfo {title} {The dripping-to-jetting transition in a
  co-axial flow of aqueous two-phase systems with low interfacial tension},}\
  }\href@noop {} {\bibfield  {journal} {\bibinfo  {journal} {RSC Adv.}\
  }\textbf {\bibinfo {volume} {7}},\ \bibinfo {pages} {3287--3292} (\bibinfo
  {year} {2017})}\BibitemShut {NoStop}%
\bibitem [{\citenamefont {Lo}\ \emph {et~al.}(2019)\citenamefont {Lo},
  \citenamefont {Liu}, \citenamefont {Mak}, \citenamefont {Xu}, \citenamefont
  {Chao}, \citenamefont {Li}, \citenamefont {Shum},\ and\ \citenamefont
  {Xu}}]{Xu2019}%
  \BibitemOpen
  \bibfield  {author} {\bibinfo {author} {\bibfnamefont {H.~Y.}\ \bibnamefont
  {Lo}}, \bibinfo {author} {\bibfnamefont {Y.}~\bibnamefont {Liu}}, \bibinfo
  {author} {\bibfnamefont {S.~Y.}\ \bibnamefont {Mak}}, \bibinfo {author}
  {\bibfnamefont {Z.}~\bibnamefont {Xu}}, \bibinfo {author} {\bibfnamefont
  {Y.}~\bibnamefont {Chao}}, \bibinfo {author} {\bibfnamefont {K.~J.}\
  \bibnamefont {Li}}, \bibinfo {author} {\bibfnamefont {H.~C.}\ \bibnamefont
  {Shum}}, \ and\ \bibinfo {author} {\bibfnamefont {L.}~\bibnamefont {Xu}},\
  }\bibfield  {title} {\enquote {\bibinfo {title} {Diffusion-dominated
  pinch-off of ultralow surface tension fluids},}\ }\href@noop {} {\bibfield
  {journal} {\bibinfo  {journal} {Phys. Rev. Lett.}\ }\textbf {\bibinfo
  {volume} {123}},\ \bibinfo {pages} {134501} (\bibinfo {year}
  {2019})}\BibitemShut {NoStop}%
\bibitem [{\citenamefont {Guillaument}\ \emph {et~al.}(2013)\citenamefont
  {Guillaument}, \citenamefont {Erriguible}, \citenamefont {Aymonier},
  \citenamefont {Marre},\ and\ \citenamefont
  {Subra-Paternault}}]{guillaument2013numerical}%
  \BibitemOpen
  \bibfield  {author} {\bibinfo {author} {\bibfnamefont {R.}~\bibnamefont
  {Guillaument}}, \bibinfo {author} {\bibfnamefont {A.}~\bibnamefont
  {Erriguible}}, \bibinfo {author} {\bibfnamefont {C.}~\bibnamefont
  {Aymonier}}, \bibinfo {author} {\bibfnamefont {S.}~\bibnamefont {Marre}}, \
  and\ \bibinfo {author} {\bibfnamefont {P.}~\bibnamefont {Subra-Paternault}},\
  }\bibfield  {title} {\enquote {\bibinfo {title} {Numerical simulation of
  dripping and jetting in supercritical fluids/liquid micro coflows},}\
  }\href@noop {} {\bibfield  {journal} {\bibinfo  {journal} {J. Supercrit.
  Fluids.}\ }\textbf {\bibinfo {volume} {81}},\ \bibinfo {pages} {15--22}
  (\bibinfo {year} {2013})}\BibitemShut {NoStop}%
\bibitem [{\citenamefont {Lei}\ and\ \citenamefont
  {Wang}(2011)}]{lei2011dripping}%
  \BibitemOpen
  \bibfield  {author} {\bibinfo {author} {\bibfnamefont {S.-L.}\ \bibnamefont
  {Lei}}\ and\ \bibinfo {author} {\bibfnamefont {X.}~\bibnamefont {Wang}},\
  }\bibfield  {title} {\enquote {\bibinfo {title} {Dripping and jetting in
  coflowing liquid streams},}\ }\href@noop {} {\bibfield  {journal} {\bibinfo
  {journal} {Adv. adapt data analysis}\ }\textbf {\bibinfo {volume} {3}},\
  \bibinfo {pages} {269--290} (\bibinfo {year} {2011})}\BibitemShut {NoStop}%
\bibitem [{\citenamefont {Shahin}\ and\ \citenamefont
  {Mortazavi}(2017)}]{shahin2017three}%
  \BibitemOpen
  \bibfield  {author} {\bibinfo {author} {\bibfnamefont {H.}~\bibnamefont
  {Shahin}}\ and\ \bibinfo {author} {\bibfnamefont {S.}~\bibnamefont
  {Mortazavi}},\ }\bibfield  {title} {\enquote {\bibinfo {title}
  {Three-dimensional simulation of microdroplet formation in a co-flowing
  immiscible fluid system using front tracking method},}\ }\href@noop {}
  {\bibfield  {journal} {\bibinfo  {journal} {J. Mol. Liq.}\ }\textbf {\bibinfo
  {volume} {243}},\ \bibinfo {pages} {737--749} (\bibinfo {year}
  {2017})}\BibitemShut {NoStop}%
\bibitem [{\citenamefont {Anderson}, \citenamefont {McFadden},\ and\
  \citenamefont {Wheeler}(1998)}]{anderson1998diffuse}%
  \BibitemOpen
  \bibfield  {author} {\bibinfo {author} {\bibfnamefont {D.~M.}\ \bibnamefont
  {Anderson}}, \bibinfo {author} {\bibfnamefont {G.~B.}\ \bibnamefont
  {McFadden}}, \ and\ \bibinfo {author} {\bibfnamefont {A.~A.}\ \bibnamefont
  {Wheeler}},\ }\bibfield  {title} {\enquote {\bibinfo {title}
  {Diffuse-interface methods in fluid mechanics},}\ }\href@noop {} {\bibfield
  {journal} {\bibinfo  {journal} {Annu. Rev. Fluid Mech.}\ }\textbf {\bibinfo
  {volume} {30}},\ \bibinfo {pages} {139--165} (\bibinfo {year}
  {1998})}\BibitemShut {NoStop}%
\bibitem [{\citenamefont {Boettinger}\ \emph {et~al.}(2002)\citenamefont
  {Boettinger}, \citenamefont {Warren}, \citenamefont {Beckermann},\ and\
  \citenamefont {Karma}}]{boettinger2002phase}%
  \BibitemOpen
  \bibfield  {author} {\bibinfo {author} {\bibfnamefont {W.~J.}\ \bibnamefont
  {Boettinger}}, \bibinfo {author} {\bibfnamefont {J.~A.}\ \bibnamefont
  {Warren}}, \bibinfo {author} {\bibfnamefont {C.}~\bibnamefont {Beckermann}},
  \ and\ \bibinfo {author} {\bibfnamefont {A.}~\bibnamefont {Karma}},\
  }\bibfield  {title} {\enquote {\bibinfo {title} {Phase-field simulation of
  solidification},}\ }\href@noop {} {\bibfield  {journal} {\bibinfo  {journal}
  {Annu. Rev. Mater. Res.}\ }\textbf {\bibinfo {volume} {32}},\ \bibinfo
  {pages} {163--194} (\bibinfo {year} {2002})}\BibitemShut {NoStop}%
\bibitem [{\citenamefont {Yue}\ \emph {et~al.}(2004)\citenamefont {Yue},
  \citenamefont {Feng}, \citenamefont {Liu},\ and\ \citenamefont
  {Shen}}]{yue2004diffuse}%
  \BibitemOpen
  \bibfield  {author} {\bibinfo {author} {\bibfnamefont {P.}~\bibnamefont
  {Yue}}, \bibinfo {author} {\bibfnamefont {J.~J.}\ \bibnamefont {Feng}},
  \bibinfo {author} {\bibfnamefont {C.}~\bibnamefont {Liu}}, \ and\ \bibinfo
  {author} {\bibfnamefont {J.}~\bibnamefont {Shen}},\ }\bibfield  {title}
  {\enquote {\bibinfo {title} {A diffuse-interface method for simulating
  two-phase flows of complex fluids},}\ }\href@noop {} {\bibfield  {journal}
  {\bibinfo  {journal} {J. Fluid Mech.}\ }\textbf {\bibinfo {volume} {515}},\
  \bibinfo {pages} {293--317} (\bibinfo {year} {2004})}\BibitemShut {NoStop}%
\bibitem [{\citenamefont {Qian}, \citenamefont {Wang},\ and\ \citenamefont
  {Sheng}(2003)}]{Qian2003}%
  \BibitemOpen
  \bibfield  {author} {\bibinfo {author} {\bibfnamefont {T.}~\bibnamefont
  {Qian}}, \bibinfo {author} {\bibfnamefont {X.~P.}\ \bibnamefont {Wang}}, \
  and\ \bibinfo {author} {\bibfnamefont {P.}~\bibnamefont {Sheng}},\ }\bibfield
   {title} {\enquote {\bibinfo {title} {Molecular scale contact line
  hydrodynamics of immiscible flows},}\ }\href@noop {} {\bibfield  {journal}
  {\bibinfo  {journal} {Phys. Rev. E}\ }\textbf {\bibinfo {volume} {68}},\
  \bibinfo {pages} {016306} (\bibinfo {year} {2003})}\BibitemShut {NoStop}%
\bibitem [{\citenamefont {Qian}, \citenamefont {Wang},\ and\ \citenamefont
  {Sheng}(2004)}]{Qian2004}%
  \BibitemOpen
  \bibfield  {author} {\bibinfo {author} {\bibfnamefont {T.}~\bibnamefont
  {Qian}}, \bibinfo {author} {\bibfnamefont {X.~P.}\ \bibnamefont {Wang}}, \
  and\ \bibinfo {author} {\bibfnamefont {P.}~\bibnamefont {Sheng}},\ }\bibfield
   {title} {\enquote {\bibinfo {title} {Power-law slip profile of the moving
  contact line in two-phase immiscible flows},}\ }\href@noop {} {\bibfield
  {journal} {\bibinfo  {journal} {Phys. Rev. Lett.}\ }\textbf {\bibinfo
  {volume} {93}},\ \bibinfo {pages} {094501} (\bibinfo {year}
  {2004})}\BibitemShut {NoStop}%
\bibitem [{\citenamefont {Yang}\ \emph {et~al.}(2006)\citenamefont {Yang},
  \citenamefont {Feng}, \citenamefont {Liu},\ and\ \citenamefont
  {Shen}}]{yang2006JCP}%
  \BibitemOpen
  \bibfield  {author} {\bibinfo {author} {\bibfnamefont {X.}~\bibnamefont
  {Yang}}, \bibinfo {author} {\bibfnamefont {J.~J.}\ \bibnamefont {Feng}},
  \bibinfo {author} {\bibfnamefont {C.}~\bibnamefont {Liu}}, \ and\ \bibinfo
  {author} {\bibfnamefont {J.}~\bibnamefont {Shen}},\ }\bibfield  {title}
  {\enquote {\bibinfo {title} {Numerical simulations of jet pinching-off and
  drop formation using an energetic variational phase-field method},}\
  }\href@noop {} {\bibfield  {journal} {\bibinfo  {journal} {J. Comput. Phys.}\
  }\textbf {\bibinfo {volume} {218}},\ \bibinfo {pages} {417--428} (\bibinfo
  {year} {2006})}\BibitemShut {NoStop}%
\bibitem [{\citenamefont {Huang}, \citenamefont {Bao},\ and\ \citenamefont
  {Qian}(2022)}]{huang2022diffuse}%
  \BibitemOpen
  \bibfield  {author} {\bibinfo {author} {\bibfnamefont {F.}~\bibnamefont
  {Huang}}, \bibinfo {author} {\bibfnamefont {W.}~\bibnamefont {Bao}}, \ and\
  \bibinfo {author} {\bibfnamefont {T.}~\bibnamefont {Qian}},\ }\bibfield
  {title} {\enquote {\bibinfo {title} {Diffuse-interface approach to
  competition between viscous flow and diffusion in pinch-off dynamics},}\
  }\href@noop {} {\bibfield  {journal} {\bibinfo  {journal} {Phys. Rev.
  Fluids}\ }\textbf {\bibinfo {volume} {7}},\ \bibinfo {pages} {094004}
  (\bibinfo {year} {2022})}\BibitemShut {NoStop}%
\bibitem [{\citenamefont {Shen}(1997)}]{Shen97}%
  \BibitemOpen
  \bibfield  {author} {\bibinfo {author} {\bibfnamefont {J.}~\bibnamefont
  {Shen}},\ }\bibfield  {title} {\enquote {\bibinfo {title} {Efficient
  spectral-{G}alerkin methods {III}: Polar and cylindrical geometries},}\
  }\href@noop {} {\bibfield  {journal} {\bibinfo  {journal} {SIAM J. Sci.
  Comput.}\ }\textbf {\bibinfo {volume} {18}},\ \bibinfo {pages} {1583--1604}
  (\bibinfo {year} {1997})}\BibitemShut {NoStop}%
\bibitem [{\citenamefont {Shen}(1992)}]{shen1992error}%
  \BibitemOpen
  \bibfield  {author} {\bibinfo {author} {\bibfnamefont {J.}~\bibnamefont
  {Shen}},\ }\bibfield  {title} {\enquote {\bibinfo {title} {On error estimates
  of projection methods for {N}avier--{S}tokes equations: first-order
  schemes},}\ }\href@noop {} {\bibfield  {journal} {\bibinfo  {journal} {SIAM
  J. Numer. Anal.}\ }\textbf {\bibinfo {volume} {29}},\ \bibinfo {pages}
  {57--77} (\bibinfo {year} {1992})}\BibitemShut {NoStop}%
\bibitem [{\citenamefont {Lopez}\ and\ \citenamefont
  {Shen}(1998)}]{lopez1998efficient}%
  \BibitemOpen
  \bibfield  {author} {\bibinfo {author} {\bibfnamefont {J.}~\bibnamefont
  {Lopez}}\ and\ \bibinfo {author} {\bibfnamefont {J.}~\bibnamefont {Shen}},\
  }\bibfield  {title} {\enquote {\bibinfo {title} {An efficient
  spectral-projection method for the {N}avier--{S}tokes equations in
  cylindrical geometries: {I}. axisymmetric cases},}\ }\href@noop {} {\bibfield
   {journal} {\bibinfo  {journal} {J. Comput. Phys.}\ }\textbf {\bibinfo
  {volume} {139}},\ \bibinfo {pages} {308--326} (\bibinfo {year}
  {1998})}\BibitemShut {NoStop}%
\bibitem [{\citenamefont {Lopez}, \citenamefont {Marques},\ and\ \citenamefont
  {Shen}(2002)}]{lopez2002efficient}%
  \BibitemOpen
  \bibfield  {author} {\bibinfo {author} {\bibfnamefont {J.}~\bibnamefont
  {Lopez}}, \bibinfo {author} {\bibfnamefont {F.}~\bibnamefont {Marques}}, \
  and\ \bibinfo {author} {\bibfnamefont {J.}~\bibnamefont {Shen}},\ }\bibfield
  {title} {\enquote {\bibinfo {title} {An efficient spectral-projection method
  for the {N}avier--{S}tokes equations in cylindrical geometries: {II}.
  three-dimensional cases},}\ }\href@noop {} {\bibfield  {journal} {\bibinfo
  {journal} {J. Comput. Phys.}\ }\textbf {\bibinfo {volume} {176}},\ \bibinfo
  {pages} {384--401} (\bibinfo {year} {2002})}\BibitemShut {NoStop}%
\bibitem [{\citenamefont {Qian}, \citenamefont {Wang},\ and\ \citenamefont
  {Sheng}(2006)}]{Qian2006}%
  \BibitemOpen
  \bibfield  {author} {\bibinfo {author} {\bibfnamefont {T.}~\bibnamefont
  {Qian}}, \bibinfo {author} {\bibfnamefont {X.~P.}\ \bibnamefont {Wang}}, \
  and\ \bibinfo {author} {\bibfnamefont {P.}~\bibnamefont {Sheng}},\ }\bibfield
   {title} {\enquote {\bibinfo {title} {A variational approach to moving
  contact line hydrodynamics},}\ }\href@noop {} {\bibfield  {journal} {\bibinfo
   {journal} {J. Fluid Mech.}\ }\textbf {\bibinfo {volume} {564}},\ \bibinfo
  {pages} {333--360} (\bibinfo {year} {2006})}\BibitemShut {NoStop}%
\bibitem [{\citenamefont {Onsager}(1931{\natexlab{a}})}]{Onsager1931a}%
  \BibitemOpen
  \bibfield  {author} {\bibinfo {author} {\bibfnamefont {L.}~\bibnamefont
  {Onsager}},\ }\bibfield  {title} {\enquote {\bibinfo {title} {Reciprocal
  relations in irreversible processes. {I}.}}\ }\href@noop {} {\bibfield
  {journal} {\bibinfo  {journal} {Phys. Rev.}\ }\textbf {\bibinfo {volume}
  {37}},\ \bibinfo {pages} {405} (\bibinfo {year}
  {1931}{\natexlab{a}})}\BibitemShut {NoStop}%
\bibitem [{\citenamefont {Onsager}(1931{\natexlab{b}})}]{onsager1931b}%
  \BibitemOpen
  \bibfield  {author} {\bibinfo {author} {\bibfnamefont {L.}~\bibnamefont
  {Onsager}},\ }\bibfield  {title} {\enquote {\bibinfo {title} {Reciprocal
  relations in irreversible processes. {II}.}}\ }\href@noop {} {\bibfield
  {journal} {\bibinfo  {journal} {Phys. Rev.}\ }\textbf {\bibinfo {volume}
  {38}},\ \bibinfo {pages} {2265} (\bibinfo {year}
  {1931}{\natexlab{b}})}\BibitemShut {NoStop}%
\bibitem [{\citenamefont {Cahn}\ and\ \citenamefont {Hilliard}(1958)}]{CH1958}%
  \BibitemOpen
  \bibfield  {author} {\bibinfo {author} {\bibfnamefont {J.~W.}\ \bibnamefont
  {Cahn}}\ and\ \bibinfo {author} {\bibfnamefont {J.~E.}\ \bibnamefont
  {Hilliard}},\ }\bibfield  {title} {\enquote {\bibinfo {title} {Free energy of
  a nonuniform system. {I}. interfacial free energy},}\ }\href@noop {}
  {\bibfield  {journal} {\bibinfo  {journal} {J. Chem. Phys.}\ }\textbf
  {\bibinfo {volume} {28}},\ \bibinfo {pages} {258--267} (\bibinfo {year}
  {1958})}\BibitemShut {NoStop}%
\bibitem [{\citenamefont {Dong}\ and\ \citenamefont {Shen}(2012)}]{dong2012}%
  \BibitemOpen
  \bibfield  {author} {\bibinfo {author} {\bibfnamefont {S.}~\bibnamefont
  {Dong}}\ and\ \bibinfo {author} {\bibfnamefont {J.}~\bibnamefont {Shen}},\
  }\bibfield  {title} {\enquote {\bibinfo {title} {A time-stepping scheme
  involving constant coefficient matrices for phase-field simulations of
  two-phase incompressible flows with large density ratios},}\ }\href@noop {}
  {\bibfield  {journal} {\bibinfo  {journal} {J. Comput. Phys.}\ }\textbf
  {\bibinfo {volume} {231}},\ \bibinfo {pages} {5788--5804} (\bibinfo {year}
  {2012})}\BibitemShut {NoStop}%
\end{thebibliography}%

\end{document}